\documentclass[twoside,leqno,twocolumn, 10pt]{article}
\usepackage[letterpaper]{geometry}
\usepackage{ltexpprt}
\usepackage{eqparbox}
\usepackage{multirow}
\usepackage{subcaption}
\usepackage{amsmath, amssymb}
\usepackage{siunitx}
\usepackage[colorinlistoftodos]{todonotes}
\usepackage{algorithm}
\usepackage[noend]{algorithmic}
\usepackage{array}
\usepackage{float}
\usepackage{verbatim}
\usepackage{xspace}
\usepackage{enumitem}
\usepackage[capitalise]{cleveref}
\usepackage[a]{esvect}
\usepackage{url}
\usepackage{tabularx}
\usepackage{multibib}
\newcites{appendix}{References}

\newcommand{\mat}[1]{\mathbf{#1}}

\newcommand{\methodname}{HYPA\xspace}

\newcommand{\mean}[1]{\langle #1 \rangle}

\begin{document}

\title{\Large HYPA: Efficient Detection of Path Anomalies in Time Series Data on Networks
\vspace{-0.35cm}}
\author{Timothy LaRock\thanks{Network Science Institute, Northeastern University}
\and Vahan Nanumyan\thanks{Chair of Systems Design, ETH Z\"urich}
\and Ingo Scholtes\thanks{Chair of Data Analytics, University of Wuppertal} \thanks{Data Analytics Group, IfI, University of Z\"urich}
\and Giona Casiraghi\footnotemark[2]
\and Tina Eliassi-Rad\footnotemark[1] \thanks{Khoury College of Computer Sciences, Northeastern University}
\and Frank Schweitzer\footnotemark[2]}

\date{\vspace{-0.85cm}}
\maketitle


\fancyfoot[R]{\scriptsize{Copyright \textcopyright\ 2020 by SIAM\\
Unauthorized reproduction of this article is prohibited}}


\begin{abstract}
\small\baselineskip=9pt The unsupervised detection of anomalies in time series data has important applications in user behavioral modeling, fraud detection, and cybersecurity.
Anomaly detection has, in fact, been extensively studied in categorical sequences. 
However, we often have access to time series data that represent \emph{paths} through networks.
Examples include transaction sequences in financial networks, click streams of users in networks of cross-referenced documents, or travel itineraries in transportation networks.
To reliably detect anomalies, we must account for the fact that such data contain a large number of independent observations of paths constrained by a graph topology.
Moreover, the heterogeneity of real systems rules out frequency-based anomaly detection techniques, which do not account for highly skewed edge and degree statistics.
To address this problem, we introduce \methodname, a novel framework for the unsupervised detection of anomalies in large corpora of variable-length temporal paths in a graph.
\methodname provides an efficient analytical method to detect paths with anomalous frequencies that result from nodes being traversed in unexpected chronological order.
\vspace{-0.3cm}
\end{abstract}

\section{Introduction}
\label{sec:intro}

Anomaly detection refers to the problem of finding ``patterns in data that do not conform to a well-defined notion of normal behavior'' \cite{chandola2009anomaly}.
The importance of anomaly detection techniques rests on the fact that anomalous patterns may carry valuable meaning.
Examples include anomalous usage or traffic patterns used to detect cyberattacks, anomalous sensor readings that may identify imminent faults in technical systems, or anomalous transaction patterns used to detect fraud and compliance violations in financial systems.
In order to assess which data represent ``anomalies'', we must define what we consider ``normal'' behavior in the particular system under study.
Given this baseline of ``normal'' behavior, we need methods to efficiently assess which patterns in the data exhibit deviations from this baseline. 
Finally, we need techniques to argue which of those observed deviant patterns are \emph{significant} given the fluctuations and randomness contained in data.

\begin{figure}[!ht]
\centering
\includegraphics[width=.75\columnwidth]{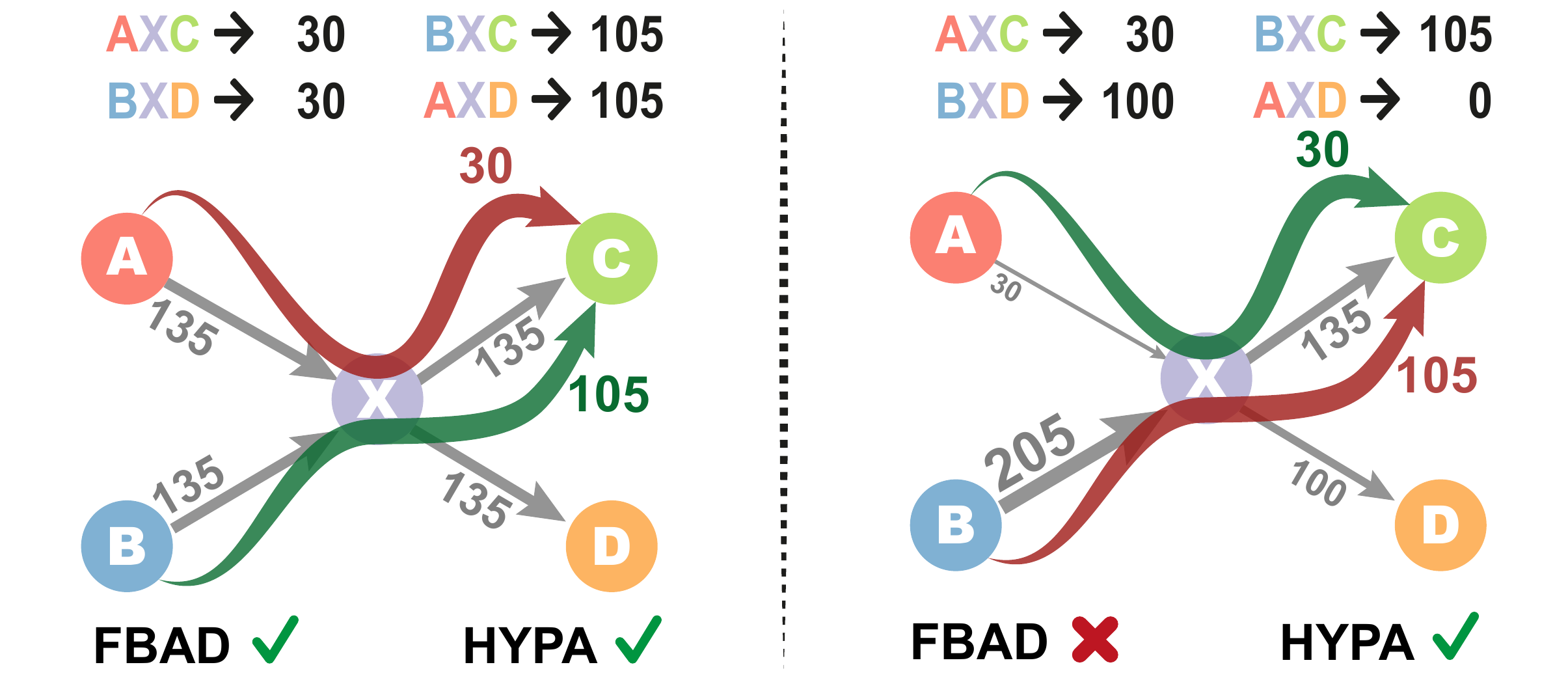}
\vspace{-0.2cm}
\caption{\small Frequency-based anomaly detection (FBAD) can be used to identify ground truth under- (red) and over-represented (green) paths in graph with homogenous edge statistics (left), but fails to identify anomalies in data with heterogeneous edge statistics (right). Our proposed method HYPA succeeds in both scenarios.}
\label{fig:motivation}
\vspace{0.15cm}
\end{figure}

Anomaly detection has been studied extensively for general categorical sequence data. However, we are often confronted with time series data capturing \emph{paths through networks}.
Such data have distinctive characteristics. Different from general categorical sequences, an underlying graph topology constrains which paths, i.e., sequences of node traversals, can possibly occur.
Moreover, the graphs in which paths are observed often exhibit strong heterogeneities, e.g., heavily skewed node degree distributions or heterogeneous edge statistics.\footnote{In this paper, we use the term \emph{heterogeneous} in reference to statistical distributions of edge and path frequencies in networks. We are \emph{not} working with \emph{heterogeneous graphs}, where nodes and edges in the same graph may have different types (e.g. \cite{liu2018subgraph}).}
These heterogeneities invalidate frequency-based anomaly detection techniques that do not account for the fact that in real systems, some paths are more likely to be observed at random than others (see Fig.~\ref{fig:motivation}).
\vspace{-0.1cm}

Closing this gap, we consider the problem of detecting \emph{anomalous paths} through graphs based on data capturing sequences of node traversals.
Our definition of \emph{anomalous paths} rests on a memoryless baseline model, which assumes that the chronological order of node traversals is determined by the graph topology and the edge traversal statistics.
We develop \methodname, an algorithm for detecting paths with unexpected temporal traversal patterns.

This problem is of practical relevance in a number of scenarios.
For example, in graphs representing transportation systems, such as passenger flights, we can study trajectories generated by passengers navigating through the network. Here anomalous paths convey information about the role of airports in routing people through the system.

Our main contributions are:
\begin{enumerate}[noitemsep,topsep=0pt,leftmargin=0.5cm,label={(\roman*)}]
    \item We introduce \emph{path anomalies}, paths through a graph that are traversed significantly more or less often than expected under a null model. We show that the problem of detecting length \(k\) path anomalies can be reduced to detecting \emph{anomalous edges} in a $k$-dimensional De Bruijn graph model of paths.

    \item We introduce \methodname, an algorithm for detecting path anomalies. 
    \methodname\ finds paths that occur significantly more or less often than expected at random, leveraging an analytically tractable statistical model of random weighted De Bruijn graphs to derive closed-form expressions for the cumulative weight distribution of paths of any length \(k\).
    \item We test \methodname\ in empirical data representing trajectories through transportation systems, validating detected anomalies with geographical information.
\end{enumerate}

The remainder of this paper is organized as follows. In the next section, we discuss related work and introduce relevant background that forms the basis of our method. In \cref{sec:method}, we formally define path anomalies, walk through an illustrative example, and introduce \methodname. In \cref{sec:data}, we validate \methodname\ in synthetic data, before applying it to analyze a dataset of passenger trips through an airport network.

\begin{figure*}[!ht]
  \centering
  \vspace{-1.5cm}
  \footnotesize
  \includegraphics[width=0.85\linewidth]{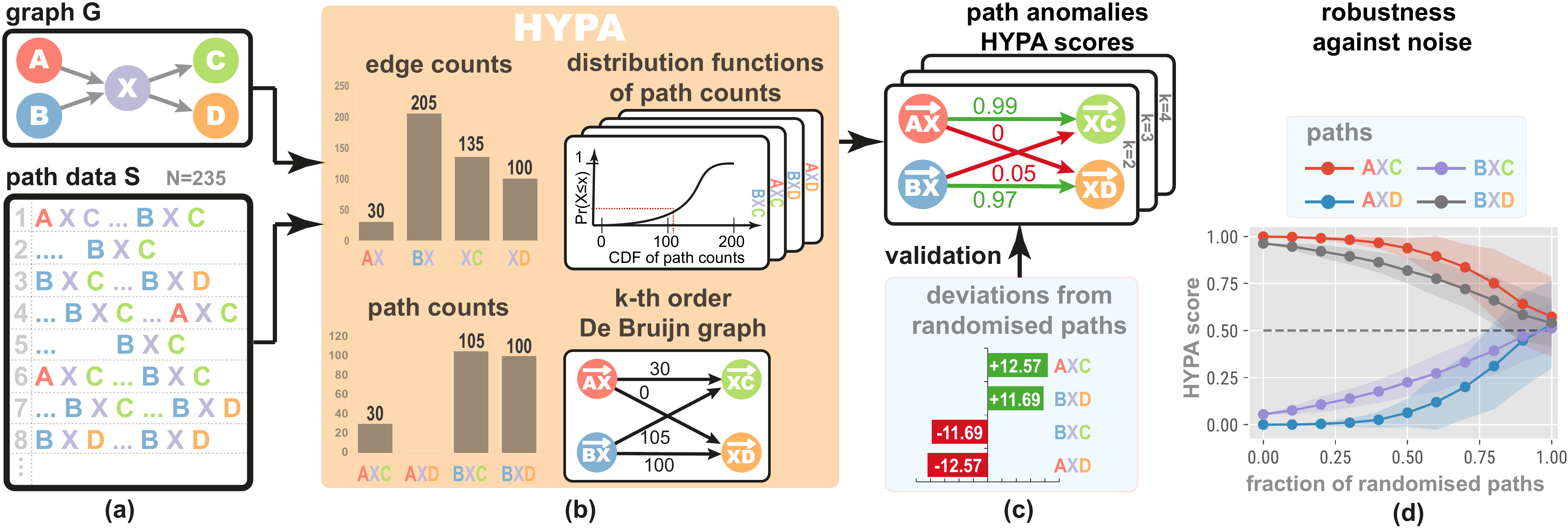}
  \vspace{-0.2cm}
   	\caption{ \small Example of path data \(\mathbf{S}\) observed in a graph \(G\) illustrates path anomaly detection with \methodname (focusing on k=2).
   Given a set of sequences traversing nodes \(A\), \(B\), \(X\), \(C\), and \(D\) in a graph (a), 
   \methodname uses higher-order De Bruijn graphs to derive closed-form expressions for the cumulative distribution function of all possible paths in the graph (b).
   \methodname computes HYPA-scores (c) that allow reliable detection of over- and under-represented paths, even in situations where the least frequent path (\(AXC\)) is over-represented, while the most frequent path (\(BXC\)) is under-represented.
   Progressive randomization of the data gradually levels HYPA-scores (d), translating to a decreasing confidence in detected anomalies.}\label{fig:toy-examples}
	\vspace{-0.16cm}
\end{figure*}

\vspace{-0.25cm}
\section{Related Work and Background}
\label{sec:background}

In this section we summarize related work on anomaly detection and sequential pattern mining and provide background on the higher-order graph models and statistical graph ensembles underlying \methodname.

\vspace{-0.25cm}
\subsection{Related Work}
\label{sec:relatedwork}

Considering the large body of research on anomaly detection in time series data~\cite{gupta2014}, and keeping in mind the focus of this paper, we limit our review to related work on (i) anomaly detection in discrete sequences, (ii) sequential pattern mining, and (iii) graph-based anomaly detection.
Since we are concerned with the unsupervised detection of path anomalies, we further exclude \mbox{(semi-)}supervised and reinforcement learning techniques.

\emph{Anomaly Detection in Sequence Data.}
Following \cite{chandola2008comparative,chandola2012}, anomaly detection techniques for discrete sequences fall into different categories that address fundamentally different application scenarios.
Sequence-based anomaly detection assumes that we are given a set \(\mathbf{S} = \{ s_1, s_2, \ldots, s_n\}\) of sequences \(s_i=(x_j)_{j=1, \ldots, l_i}\) over a discrete alphabet \(\Sigma\), possibly with variable lengths \(l_i\).
Anomalous instances \(s_i\) in \(\mathbf{S}\) are then detected. 
For example, each sequence may be assigned an anomaly score, then ranked from most to least anomalous by the magnitude of this score.
Different approaches have been used to establish a random baseline against which sequences are defined as ``anomalous''.
Some methods have used (hidden) Markov chain models, e.g., to detect (groups of) sequences which show significant differences in terms of state transition probabilities~\cite{smyth1997clustering,lane2003empirical,lemmerich2016mining,atzmueller2016detecting}.
Other methods use nearest-neighbours algorithms \cite{melnyk2016vector} or distance measures~\cite{tonnelier2018anomaly} to quantify how any given sequence \(s_j\) differs from other instances in \(\mathbf{S}\).
Adopting a collective definition of anomalies~\cite{chandola2009anomaly}, a third class of methods is based on hypothesis testing techniques to detect outliers in the distribution of features of sequences~\cite{tajer2014outlying,laxhammar2014online,atzmueller2016sequential}.
Our problem setting is different because we are interested in discovering patterns in an underlying network structure, not in marking individual sequences as anomalous.

\emph{Sequential pattern mining.}
A common feature of the methods outlined above is that they focus on anomalies at the level of a whole sequence \(s_i\) within \(\mathbf{S}\).
Addressing a different problem, some methods instead attempt to find anomalous \emph{patterns} or \emph{subsequences} within a long sequence \(S = \left(x_i\right)_{i=1, \ldots, n}\)~\cite{chandola2012}.
This is called sequential pattern mining, where the goal is to develop algorithms that quickly find the most frequent subsequences in large sequence data~\cite{agrawal1995mining, jianpei2001prefixspan, Sayed2004,servan2018prosecco}.
Some work addresses this problem based on statistical methods, e.g., using Markov modeling techniques~\cite{gwadera2005markov, sadoddin2016finding, zhou2016bi,cadez2000visualization,walk2014,peixoto2017modelling, gwadera2010ranking}, hypothesis testing~\cite{singer2015,becker2017mixedtrails, tonon2019permutation}, or information-theoretic methods to detect ``surprising'' subsequences~\cite{keogh2007compression,chakrabarti1998mining,bertens2016keeping}.
Applications include the detection of common patterns in user trajectories~\cite{walk2014,Rosvall2014}, testing hypotheses about generative processes of trajectory data~\cite{singer2015}, or finding clusters in sequence data~\cite{cadez2000visualization, peixoto2017modelling}. 

\emph{Temporal Anomaly Detection in Graphs.}
The problem motivating our method is different from those described so far, mainly due to the fact that these methods make no assumptions about the relational structure of the data, while we study sequential data capturing \emph{paths in a (weighted and directed) graph topology}.
This aligns our work more closely with anomaly detection techniques for temporal graph data that have been developed in the graph mining community \cite{Noble2003,akoglu2015graph}.
As summarized in~\cite{akoglu2015graph}, temporal anomaly detection discovers change events~\cite{akoglu2010event} or cluster structures in evolving graphs~\cite{boden2012tracing, bogdanov2011mining, peixoto2017modelling}. 

Different from these problems, our method uses a set \(\mathbf{S}\) of sequences to identify paths through a graph that are traversed more or less often than expected.
Hence, rather than making statements about anomalous instances in \(\mathbf{S}\), we use collective statistical information in \(\mathbf{S}\) to identify paths through the graph that are traversed with \emph{anomalous frequencies}.

\vspace{-0.25cm}
\subsection{Background}
\label{subsec:background}

In this section, we provide definitions necessary to the formulation of path anomaly detection and our solution. 

We reduce the problem of detecting anomalous \emph{paths} in a (first-order) graph to detection of anomalous \emph{edges} in \emph{higher-order} graph models that resemble De Bruijn graphs~\cite{DeBruijn1946}.
Similar to \cite{Scholtes2017}, we define a higher-order De Bruijn graph model of paths as:
\begin{Definition}[\small \bf \(k\)-th order De Bruijn graph model]
\label{def:debruijn}
For a given graph \(G=(V,E)\) and $k \in \mathbb{N}$ we define a \(k\)-th order De Bruijn graph of paths in \(G\) as a graph $G^k=(V^k, E^k)$, where (i) each node \(\vv{v}:=\vv{v_0v_1\ldots v_{k-1}} \in V^k\) is a path of length\footnote{We assume path length is the number of edges traversed.} \(k-1\) in \(G\), and (ii) \((\vv{v},\vv{w}) \in E^k\) iff \(v_{i+1} = w_{i}\) for \(i=0, \ldots, k-2\).
\end{Definition}
This definition has several implications.
First, any two nodes \(\vv{v}\) and \(\vv{w}\) connected by an edge in a \(k\)-th order graph $G^k$ represent 2 paths of length \(k-1\) that overlap in exactly \(k-1\) out of $k$ nodes.
Since paths in a graph are transitive, each edge \((\vv{v},\vv{w})\) in $G^k$ represents a path of length \(k\) in graph \(G\).
This implies that the graph \(G\) itself is a first-order De Bruijn graph of paths of length one (i.e., edges) in \(G\).
We can see De Bruijn graphs as a generalization of standard, first-order graphs to higher-order models of paths of length $k$, where any path of length $q$ in $G^k$ translates to a path of length $k+q-1$ in $G$.
We iteratively construct De Bruijn graph models of order \(k\) by means of a line graph transformation on the \(k-1\)st order model.

This representation is powerful because it allows us to encode the frequencies of paths of length \(k\) through a first-order graph to the weights of edges in a \(k\)-th order De Bruijn graph.
This can be seen in the illustration of a De Bruijn graph with order \(k=2\) in \cref{fig:toy-examples}, where nodes represent paths of length \(k-1=1\) that overlap in \(k-1=1\) nodes (i.e., edges in \(G\)), while edges represent all paths of length \(k=2\).

This projection of paths allows us to reduce the problem of detecting paths of length \(k\) that exhibit anomalous frequencies to the problem of detecting anomalous edge weights in a \(k\)-th order De Bruijn graph.
To understand which edge weights exhibit ``anomalies'', we need a null model that provides a baseline against which we compare the observed weights.
For this comparison, we need to generate randomized configurations of the path data that selectively destroy only the patterns that we are interested in while preserving all other statistics.
Since we can project the path data to the edges of a directed and weighted graph, we can address this problem by employing \emph{statistical graph ensembles}, which randomize certain aspects of a graph (i.e., the weights of edges or the topology itself) while preserving other characteristics.
Examples include models that randomize the topology of a graph while preserving the (expected) number of edges~\cite{gilbert1959random}, as well as combinatorial models that preserve the degrees of nodes~\cite{Molloy1995}.

An analytically tractable formulation of such a model for directed and weighted graphs was recently proposed in~\cite{casiraghi2018generalised}.
It treats the random generation of weighted graphs as an urn problem, where random edges are drawn without replacement from a population of multi-edges connecting different pairs of nodes. 
Through this formulation, the probability of generating edges with specific weights can be calculated based on the multivariate hypergeometric distribution.
This formulation can be used to detect anomalous edges in social networks~\cite{Casiraghi2017}.
However, no analytically tractable null models have been proposed that account for the distinctive characteristics of De Bruijn graphs, i.e., the fact that a directed edge between two nodes in a \(k\)-th order De Bruijn graph can only exist if the corresponding path exists in the underlying graph.
Closing this gap, we develop a method to detect path anomalies based on statistical ensembles of \(k\)-th order De Bruijn graphs.

\vspace{-0.3cm}
\section{Higher-order Hyper-geometric path anomaly detection}
\label{sec:method}
We now define the problem of \emph{path anomaly detection}, illustrate it in an example, and propose our solution.
\vspace{-3pt}
\begin{Definition}[\bf Path Anomaly Detection]
\label{def:anomaly}
Let \(G=(V,E)\) be a directed graph and \(\mathbf{S}\) a set of \(n\) sequences $s_i$, where each sequence \(s_i = v_0v_1\ldots v_{l_i}\) is a \emph{path} of arbitrary length \(l_i\) in \(G\), i.e. \(v_j \in V\) for \mbox{\(j \in [0, \ldots, l_i]\)} and \((v_j,v_{j+1}) \in E\) for \(j \in [0, \ldots, l_i-1]\).
For \(k>1\), identify all paths \(\vv{p}=\vv{v_0\ldots v_k}\) of length \(k\) in \(G\) whose frequencies (including as subpaths) in \(\mathbf{S}\) significantly deviate from the frequencies expected in a \emph{\((k-1)\)-order} model of paths in \(G\).
\end{Definition}
\vspace{-3pt}
Unlike sequence-based anomaly detection~\cite{chandola2012}, we are not interested in assigning an \emph{anomaly score} to each sequence in \(\mathbf{S}\).
Instead we use the instances in \(\mathbf{S}\) to identify paths through the graph that exhibit anomalous frequencies compared to a null model, discovering the paths in $G$ that are traversed in unexpected ways given the underlying weighted network structure.
To complete our definition of \emph{anomalies}, we define a generative \emph{null model} for paths that builds on definition \ref{def:debruijn}, which we use to establish a baseline against which we detect anomalies.
\vspace{-3pt}
\begin{Definition}[\bf \(k\)-th order model of paths]
\label{def:nullmodel}
For a graph \(G\) let \(G^k=(V^k,E^k)\) be a \(k\)-th order De Bruijn graph of paths in \(G\) (cf. Def.~\ref{def:debruijn}).
For each edge \(e:=(\vv{v_0\ldots v_{k-1}}, \vv{v_1\ldots v_{k}}) \in E^k\) let the weight \(f(e)\) be the frequency of subpath \(\vv{v_0\ldots v_{k}}\) in \(\mathbf{S}\).
Let \(\mathbf{T}^k\) be the transition matrix of an edge-weighted random walk on \(G^k\), i.e., \(\mathbf{T}^k_{\vv{v}\vv{w}} := \frac{f(\vv{v},\vv{w})}{\sum_{\vv{x}\in V^k}f(\vv{v},\vv{x})}\).
For a path \(\vv{p}=\vv{v_0v_1\ldots v_l}\) with \(l \geq k\) the \emph{\(k\)-th order model of paths} generates \(\vv{p}\) with probability \(\prod_{i=k}^{l} \mathbf{T}^k_{\vv{v_{i-k}\ldots v_{i-1}}\vv{v_{i-k+1}\ldots v_{i}}}\).
\end{Definition}
\vspace{-3pt}
This model generates paths of length $l$ by performing $l-k+1$ random walk steps in a $k$-th order De Bruijn graph.
We can use the model to generate random paths of length \(l \geq k\) that respect (i) the topology of the underlying graph \(G\), and (ii) the frequencies of paths of length \(k\) observed in \(\mathbf{S}\).

Our definition of path anomalies of length \(k\) is based on a null model of order \(k-1\).
For \(k=2\), the null model of order \((k-1)=1\) is simply an edge-weighted random walk on the graph \(G\). 
In this case, the sequence of nodes traversed by paths is \emph{Markovian}, i.e., the node \(v_{i+1}\) on a path only depends on the current node \(v_{i}\) and the graph topology.
Apart from the topology, the model accounts for the frequencies at which paths in \(\mathbf{S}\) traverse edges in \(G\).
That is, if an edge \((b,c)\) is traversed more often than \((b,d)\) we expect path \(\vv{abc}\) to occur more often than \(\vv{abd}\).
For \(k > 2\), the null model corresponds to an edge-weighted random walk on a De Bruijn graph of order \((k-1)>1\), where weighted edges capture the frequencies of subpaths of length \(k-1\) in $\mathbf{S}$.
This approach to generating a null model is key to disentangling path anomalies that unfold at different lengths:
For any given length \(k\), we can exclusively detect those path anomalies that do not trivially result from anomalous path frequencies at shorter lengths.
In other words, to answer the question whether a \emph{specific} path \(\vv{abcd}\) of length \(k=3\) is observed more or less often than expected, we discount for any anomalies of shorter paths \(\vv{abc}\) and \(\vv{bcd}\) contained in \(\vv{abcd}\).

\vspace{-0.3cm}
\subsection{Illustrative Example}%
\label{sec:toy-example}
A simple example to illustrate the path anomaly detection problem for \(k=2\) is shown in \cref{fig:toy-examples}, which gives a high level overview of \methodname. 
Limiting our focus to paths that traverse nodes \(A\), \(B\), \(X\), \(C\), and \(D\), we consider a set \(\mathbf{S}\) that contains 235 (sub)paths of length two.
We observe strong heterogeneities in the path frequencies, where the most frequent path \(\vv{BXC}\) occurs 105 times, while the least frequent observed path \(\vv{AXC}\) occurs only 30 times.

Assume we want to detect for which paths of length \(k=2\) the frequencies deviate from the expectation in a first-order null model.
If all paths were expected to occur with similar frequency (e.g. if observed frequency was drawn from a normal distribution), we could directly answer this question based on the observed distribution of path frequencies (cf. Fig.~\ref{fig:motivation}).
Such an approach would trivially detect that path \(\vv{AXC}\) occurs more often than expected while path \(\vv{BXC}\) occurs less often than expected.
However, the edge frequencies in our toy example show strong heterogeneities; for example, edge \((B,X)\) is traversed about seven times more often than edge \((A,X)\).
If we account for this heterogeneity of edges (i.e., paths of length \(k-1=1\)), the question of which paths of length \(k=2\) exhibit statistically significant deviations becomes non-trivial.
In particular, the same observed frequency could be``normal'' (i.e., expected) for one path, a significant over-representation for another, and an under-representation for a third.
Whether the frequency of a path is anomalous based on definition \ref{def:anomaly} can not be determined by direct comparison with the overall frequency distribution alone.

We can address this problem by randomizing the data using random walk simulations on the first-order model and preserving the distribution of path lengths. 
We then count the average frequency of each path of length 2 across many simulations.
A comparison of observed vs. average frequencies of paths then indicates which paths exhibit deviations from the random baseline.
In \cref{fig:toy-examples}c, we report the average of 100 such simulation runs, which indicate that paths \(\vv{AXC}\) and \(\vv{BXD}\) occur \emph{more} often than expected, while paths \(\vv{BXC}\) and \(\vv{AXD}\) occur \emph{less} often than expected.
This simple example highlights an important problem: due to the heterogeneous frequency of edges, paths that occur with the smallest frequency (\(\vv{AXC}\)) can be over-represented, while paths that occur with the highest frequency (\(\vv{BXD}\)) can be under-represented.

This observation rules out collective anomaly detection techniques that assess anomalies based on a \emph{single} frequency distribution. 
We must instead consider the joint distribution of frequencies under the null model for each possible path and each length \(k\) separately. 
While a simulation-based approach is possible in principle, the combinatorial growth of the required computational effort for large systems is prohibitive.
Moreover, such simulations leave open the question of whether the observed deviations in the data indicate a significant pattern or are likely due to chance.

Projecting paths of length \(k\) onto edges in a \(k\)-dimensional De Bruijn graph, we use closed-form expressions for the cumulative distribution function of path frequencies under the (\(k-1\))-order null model for each path individually (see \cref{fig:toy-examples}b).
This enables us to analytically calculate \emph{HYPA-scores}, which, for each path \(\vv{p}\), represent the likelihood that a null model generates realizations where frequencies of \(\vv{p}\) are larger (or smaller) than in the data.
The calculated scores can then be used to detect path anomalies at various levels of significance without expensive simulations.

\vspace{-0.25cm}
\subsection{Hypergeometric Ensemble of Higher-Order De Bruijn Graphs}
\label{sec:method:hypa}

We now introduce the details of \emph{higher-order hypergeometric path anomaly detection} (\methodname), the main contribution of our work.

\paragraph{Mapping of null model to ensemble of $k$-th order De Bruijn graphs} In the illustrative example, we showed that assessing whether a path of length \(k\) exhibits anomalous frequencies requires considering the distribution of frequencies under a null model \emph{for each path separately}.
The key idea of \methodname is to map the difficult problem of finding the frequency distributions of paths of length \(k\) under a null model to the simpler problem of finding the edge weight distribution in a null model for \(k\)-th order De Bruijn graphs.
For this, we remember that the weights on the edges in a \(k\)-th order De Bruijn graph can exactly represent the frequencies of paths of length \(k\) observed in a dataset (cf. definition \ref{def:nullmodel}).
We are thus interested in identifying which of these weights are anomalous compared to the baseline given by a $(k-1)$-order null model of paths.
In each realization generated by such a $(k-1)$-order model, frequencies of paths of length $k-1$ are fixed, while the frequency of each path of length $k$ follows a different distribution that depends on the null model.
We can map each random realization to a different weighted $k$-th order De Bruin graph, obtaining a \emph{statistical ensemble of $k$-th order De Bruijn graphs} whose probabilities are given by the null model.
Since the frequencies of paths of length $k-1$ are fixed, the \emph{total} out-degree $f_{\vv{v}}^{\text{out}} = \sum_{\vv{x}}f(\vv{v},\vv{x})$ and the \emph{total} in-degree \(f_{\vv{v}}^{\text{in}} = \sum_{\vv{x}}f(\vv{x},\vv{v})\) for each node $\vv{v}$ is the same across all realizations in this ensemble.
However, De Bruijn graphs that correspond to different random realizations differ in terms of the exact edge weights $f(\vv{v},\vv{w})$, which represent frequencies of paths of length $k$. 

\paragraph{Distribution of edge weights in random $k$-th order graphs}
This mapping allows us to compute frequency distributions for individual paths of length \(k\), conditional on the frequencies of paths of length $k-1$, based on a random model for \(k\)-th order De Bruijn graphs that preserves the total in- and out-degrees of all nodes while randomly shuffling the weights of edges.
We can formalize the model as a stochastic process that randomly draws \(m\) edges, where \(m\) is the sum of all \(k\)-th order edge weights (the total number of paths of length $k$ observed in the data).
Different from simple random graph models, in this sampling process we must account for the fact that different edges in a $k$-th order De Bruijn graph have different probabilities to be drawn.
Specifically, we are more likely to generate edges between pairs of nodes with a high in- and out-degree.
In our null model of paths, this translates to the fact that a path of length $k$ is more likely to occur if it continues a frequently occurring path of length $k-1$.
We capture the fact that different edges in a $k$-th order De Bruijn graph occur with different probabilities using a matrix \(\Xi\), where each entry corresponds to one possible pair of higher-order nodes that can be connected by an edge, and the value of the entry denotes how many times this pair of nodes can possibly be drawn.
We thus obtain a sampling procedure that can be described by the multivariate hypergeometric distribution.

Since we consider \(k\)-th order De Bruijn graphs we must additionally account for the fact only pairs of higher-order nodes representing paths overlapping in \(k-1\) first-order nodes can be connected (cf. Def.~\ref{def:debruijn}).
When sampling from the multivariate hypergeometric distribution, we avoid drawing such pairs by setting their corresponding entry in $\Xi$ to 0.
This modification introduces the complication that weighted degrees are no longer guaranteed to be preserved, which violates the constraint that the frequency of paths of length $k-1$ is fixed.
We overcome this with an optimization approach (\cref{alg:fitxi} in \cref{app:fitXi}\footnote{Appendix A is available in the online version \cite{larock2019detecting}.}) to redistribute values of the \(\Xi\) matrix that were substituted by zeroes across the rest of the matrix, such that the weighted degrees \(f_{\vv{v}}^{\text{out}}\) and \(f_{\vv{v}}^{\text{in}}\) of the \(k\) order nodes \(\vv{v}\) are preserved.

\paragraph{\methodname Algorithm}
The random De Bruijn graph model of order $k$ introduced above is the basis for the \methodname algorithm to detect path anomalies (pseudocode in \cref{app:HYPA}, \cref{alg:hypa_algo}).
In particular, we argued that the distribution of edge weights in the statistical ensemble of random realizations are jointly described by a multivariate hypergeometric distribution.
We use the marginals of this distribution to calculate the distribution of edge weights for each edge as:
\begin{equation}
	\label{eq:dhype-ij}
	\resizebox{\linewidth}{!}{$\Pr\left(X_{\vv{v}\vv{w}} = f(\vv{v}, \vv{w})\right) = \dbinom{\sum_{ij} \Xi_{ij}}{m}^{-1} \dbinom{\Xi_{vw}}{f(\vv{v}, \vv{w})} \dbinom{\sum_{ij} \Xi_{ij} - \Xi_{vw}}{m - f(\vv{v}, \vv{w})},$}
\end{equation}
where \(m = \sum_{v} f^{\text{out}}_v = \sum_{v} f^{\text{in}}_v \) is the sum of all weights in the graph and $X_{\vv{v}\vv{w}}$ is a random variable assuming the weight of edge $(\vv{v},\vv{w})$ in a random realization of a $k$-th order De Bruin graph.
We use these marginal distributions to define the \(\text{\methodname}^{(k)}\) score for a path \({\vv{vw}}\) of length \(k\) with observed frequency \(f(\vv{v},\vv{w})\) as the cumulative distribution corresponding to \cref{eq:dhype-ij}: 
\vspace{-0.2cm}
\begin{equation}
	\label{eq:hype-ij}
	\text{\methodname}^{(k)}(\vv{v},\vv{w}) := \Pr\left(X_{\vv{v}\vv{w}} \leq f(\vv{v},\vv{w})\right) 
	\vspace{-0.2cm}
\end{equation}
Since the \(\text{\methodname}^{(k)}\) score is a probability, it assumes values in $\left[0, 1\right]$. 
Paths whose \(\text{\methodname}^{(k)}\) scores are close to 0 are likely to be under-represented compared to the random baseline.
That is, the probability to obtain at random a frequency for this path that is lower or equal to the frequency in the data is small.
On the other hand, a path whose \(\text{\methodname}^{(k)}\) score is close to 1 is likely to be over-represented, meaning the frequency obtained at random for that path is likely to be smaller than the one observed in the data.
A path that has a \(\text{\methodname}^{(k)}\) score of 0.5 is equally likely to be observed with a higher or lower frequency at random, showing the least indication of an anomaly.
Anomalous paths are determined by setting a discrimination threshold \(\alpha\ \in (0,1]\) and classifying as under-represented any path \((\vv{v},\vv{w})\) with \(\text{\methodname}^{(k)}(\vv{v},\vv{w}) < \alpha\) and as over-represented any path \((\vv{v},\vv{w})\) with \(\text{\methodname}^{(k)}(\vv{v},\vv{w}) \geq 1-\alpha\). 

\paragraph{Computational Complexity}
\label{sec:complexity}
The asymptotic runtime of \methodname\ is $\mathcal{O}\left(N+\Delta^k(G)\right)$, where $N$ is the size of $\mathbf{S}$, $\Delta^k(G)$ is the number of edges in a $k$-th order De Bruijn graph model $G^k$ of paths in $G$. An upper bound on $\Delta^k(G)$ is proved in \cref{sec:proof-complexity}.
The implication of this bound is that for sparse real-world graphs, moderate values of $k$, and above a sufficiently large value of $N$, \emph{our method scales linearly with the size of the data}.

\vspace{-0.25cm}
\section{Experiments}
\label{sec:data}
In this section we show that we can use the scores calculated by \methodname to detect paths with anomalous frequencies.\footnote{An implementation is available at \url{github.com/tlarock/hypa}.}
We note that the datasets we consider do not come with anomaly labels, and that we do not expect our notion of anomalous paths to correspond directly with any existing labeled data, since we have defined the anomalies we are detecting within a specific mathematical frame, rather than as deviation from a domain dependent ``normal''. With this in mind, we apply \methodname\ to synthetic data with known anomalies and empirical data representing trajectories through a transportation network. We show that the under- and over-represented paths detected without supervision fall into classes that can be validated using semantic and geographic information.

\vspace{-0.25cm}
\subsection{Baseline Method}
\label{sec:baseline}
In the below experiments, we compare \methodname to a simple frequency-based anomaly detection (FBAD) of our own design. We note that despite similar problem settings, the methods for hypothesis testing on human trails presented in \cite{singer2015, becker2017mixedtrails} are not directly comparable with our work because the output is Bayesian evidence for a hypothesis on an entire dataset (a single number), whereas we are interested in edge-level analysis. Further, we did not compare with a method like \cite{sadoddin2016finding} because, while based on detecting significant deviations from a Markov chain model, this method assumes that the data is given as one long sequence and detects anomalous subsequences, which does not correspond to any of the datasets we analyze here. Finally, a recently proposed method identifies significant sequential patterns via a permutation strategy with a Monte Carlo estimation procedure \cite{tonon2019permutation}. Despite similarity in purpose, the method has limited utility in our setting for two main reasons. First, it relies on the PrefixSpan algorithm \cite{jianpei2001prefixspan} to mine the sequences for relatively frequent patterns. As can be noted from \cref{fig:motivation}, low-frequency patterns can be path anomalies. Second, it does not incorporate constraints on the possible sequences, instead sampling from a uniform space of all possible permutations. For these reasons, we were unable to make a fair comparison and do not report results.

The baseline method FBAD computes the average $\mu$ and standard deviation $\sigma$ of path counts and employs a user-defined threshold $\alpha$ to detect over- and under-represented paths. FBAD implicitly assumes that the distribution of edge weights is normal and thus paths should be considered anomalous if they are outliers with respect to this distribution.
In particular, a path is labeled as over-represented if its frequency exceeds $\mu + \sigma\alpha$, and as under-represented if its frequency is smaller than $\mu - \sigma\alpha$. 
More details on FBAD are available in \cref{app:naive} and \cref{alg:naive}. 

\vspace{-0.25cm}
\subsection{Synthetic Data}
\label{sec:synthetic}
\begin{figure}[ht]
    \centering
    \vspace{-0.3cm}
    \includegraphics[width=.49\linewidth]{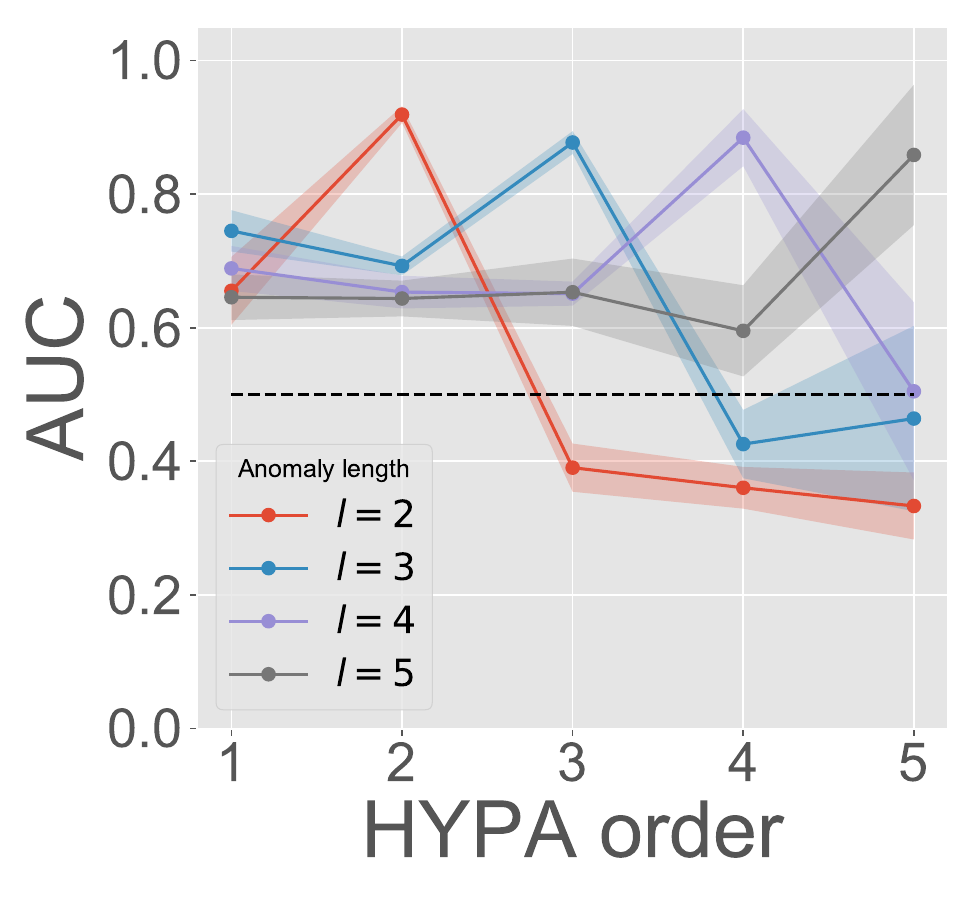}%
    \hfill
	\includegraphics[width=.49\linewidth]{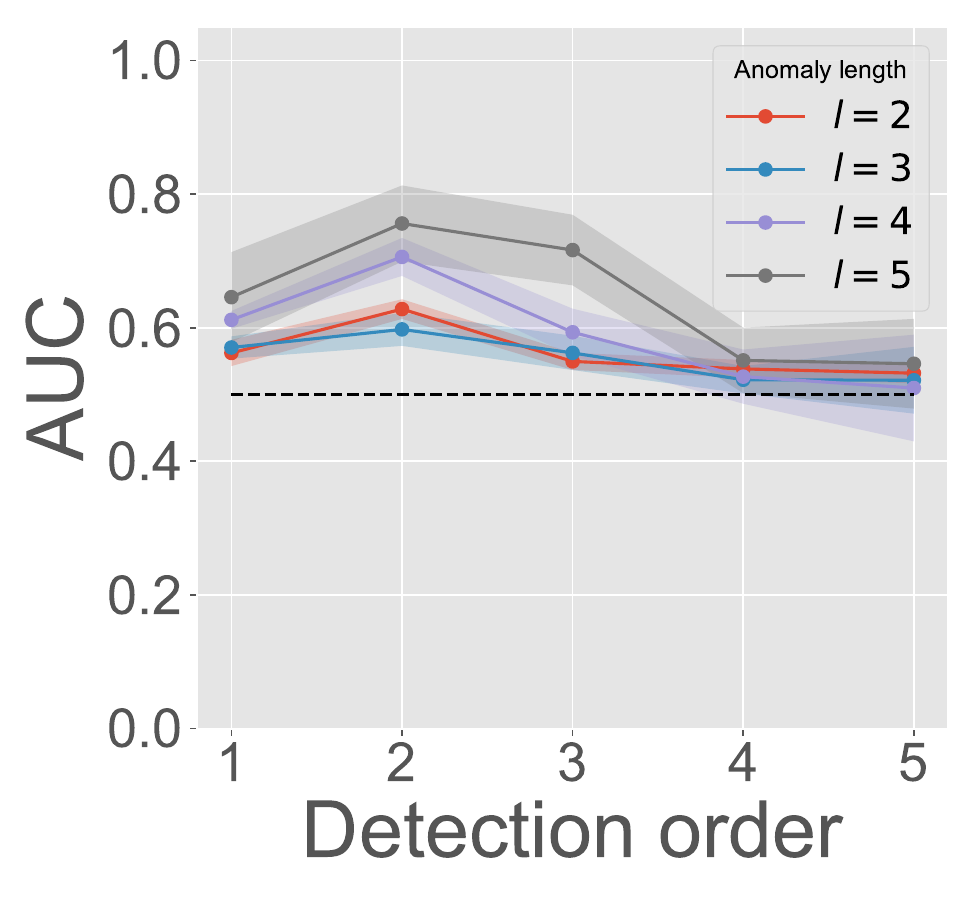}
    \caption{\small \(\text{\methodname}^{(k)}\) detects injected path anomalies at the correct length with high accuracy. 
    Each curve corresponds to one length \(l\) of generated anomalous paths, and represents the performance of classifying the anomalous paths using \methodname (left) or the naive FBAD method (right) applied at increasing orders \(k\). 
    HYPA detects the exact generated anomalies, i.e., performs highly at \(k=l\).
    FBAD only performs relatively well in detecting short sub-paths (e.g., \(k=2\)) of longer anomalies (e.g., \(l=5\)).
    Averages and standard errors are over 10 independent experiments.
    }
    \label{fig:auroc}
    \vspace{-0.2cm}
\end{figure}
We validate \methodname\ using a stochastic model that generates synthetic datasets of paths with varying lengths, in which a known set of paths with given length $l$ exhibit anomalous frequencies.
Adopting the well-known Erd\H{o}s-R\'enyi model~\cite{erdos1960evolution}, our model generates paths in a random directed graph $G$ with $n$ nodes, where pairs of nodes are connected with probability $p$.
Following definition \ref{def:nullmodel}, the random model generates paths based on an edge-weighted random walk in a $k$-th order De Bruijn graph of paths in the random graph $G$.
By selectively changing transition probabilities in $\mathbf{T}^l$ (cf. definition \ref{def:nullmodel}), we introduce anomalous frequencies for a known set of paths at length $l$.
Since all paths longer than $l$ are generated by a (Markovian) random walk on a De Bruijn graph with order $l$, these paths will not exhibit anomalous frequencies beyond those expected from the anomalous frequencies of paths of length $l$.
For details of the random path construction, see  \cref{alg:synthetic-model,alg:synthetic-walk} in \cref{app:synthetic}. 
In the following we report results for graphs with $n=50$ nodes and an edge probability of $p=0.05$ (our conclusions do not depend on these parameters).

We test whether \methodname detects anomalous path frequencies (i) with high accuracy, and (ii) at the correct length $l$ introduced by our model.
To this end, we calculate the performance of \methodname in a binary classification experiment, categorizing path frequencies as anomalous based on variable discrimination thresholds $\alpha$ for the \(\text{\methodname}^{(k)}\) scores at different orders $k$.
For each threshold $\alpha$, we compute the true and false positive rates of detected anomalies w.r.t. the known ground truth and obtain a receiver operating characteristic (ROC) curve for which we can calculate the area under the curve (AUC).
We repeat this experiment 10 times for each combination of anomalous path length \(l \in [2,5]\) and detection order $k \in [1,5]$.
Each curve in \cref{fig:auroc} presents the mean and the standard deviation of the \(AUC\) for anomalies detected at varying orders $k$, for a given anomaly length $l$.
For \(k\neq l\), we use as ground-truth the paths of length \(k\) that either include or are included in an anomalous path of length \(l\) generated by the synthetic model.
For each \(l\) we observe that \methodname with the ``correct'' order \(k=l\) is able to identify ground truth anomalies with high accuracy (\(AUC \approx 0.9\), left plot), while the baseline FBAD method (\(\sigma=2\)) is unable to detect path anomalies with high accuracy at any order, regardless of the order used for detection (max \(AUC \approx 0.78\), right).

\paragraph{Efficiency and Balance in Flight Itineraries}
We analyze an empirical dataset of paths taken through a transportation system using \methodname. 
\emph{Flights} comprise 5\% of all travel itineraries of passengers flying in the US in the first quarter of 2018.\footnote{Data from US Bureau of Transportation Statistics TransStats \url{http://www.transtats.bts.gov/Tables.asp?DB_ID=125}.} 
Characteristics of the dataset are presented in \cref{tab:data-stats} (\cref{app:data}).

Our first hypothesis is that return flights ($ABA$) are significantly over-represented, since passengers often leave from and return to the same airport.
We first compute \methodname scores for $k=2$, then separate return from non-return flights and compute the fraction of over-represented paths in each category for varying discrimination thresholds $\alpha$.
The results in \cref{tab:flights-returns} support the hypothesis that return flights are strongly over-represented compared to the null model.

\vspace{0.3cm}
\begin{table}[hbt]
    \caption{\small Fractions of over-represented paths of length 2 between airports for return flights (5840 unique paths) and non-return flights (409254 unique paths) at different discrimination thresholds $\alpha$. \vspace{-0.25cm}}
    \label{tab:flights-returns}
    \resizebox{0.99\linewidth}{!}{
    \begin{tabular}{rrrrrr}
        \(\alpha\)  & 0.05    & 0.01    & 0.001   & 0.0001  & 0.00001 \\
        \hline
        \textbf{Return} & 0.915	& 0.851	& 0.760	& 0.688	& 0.628	\\
        \textbf{Non-return} & 0.340 & 0.130 & 0.023 & 0.004 & 0.001 
    \end{tabular}
   	}
\end{table}
\vspace{-0.4cm}

\begin{figure}[ht]
	\vspace{-0.2cm}
    \centering
    \includegraphics[width=.33\linewidth]{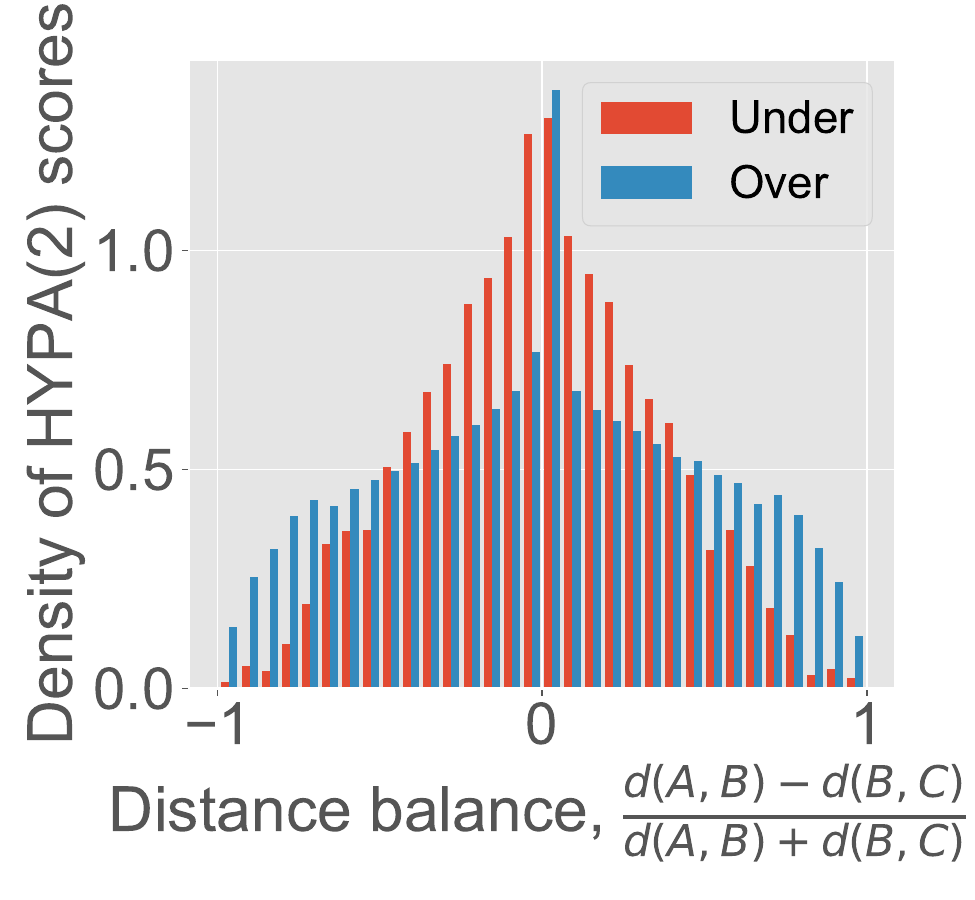}%
    \includegraphics[width=.33\linewidth]{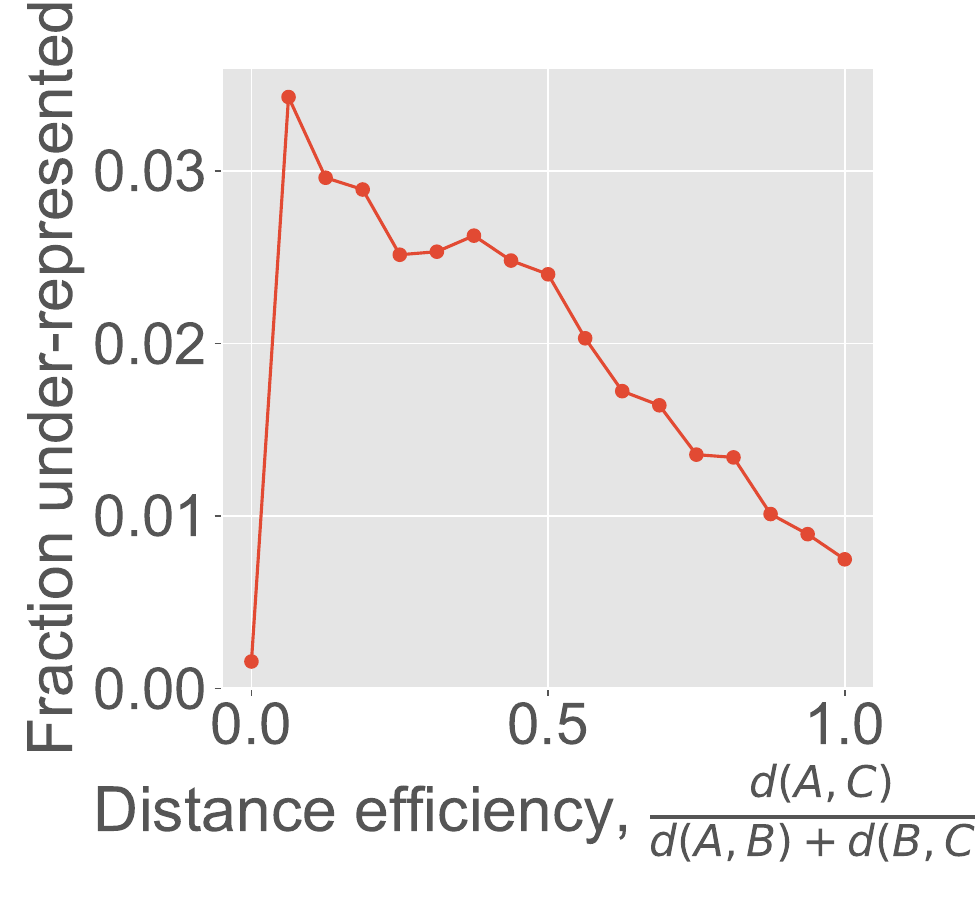}
    \includegraphics[width=.33\linewidth]{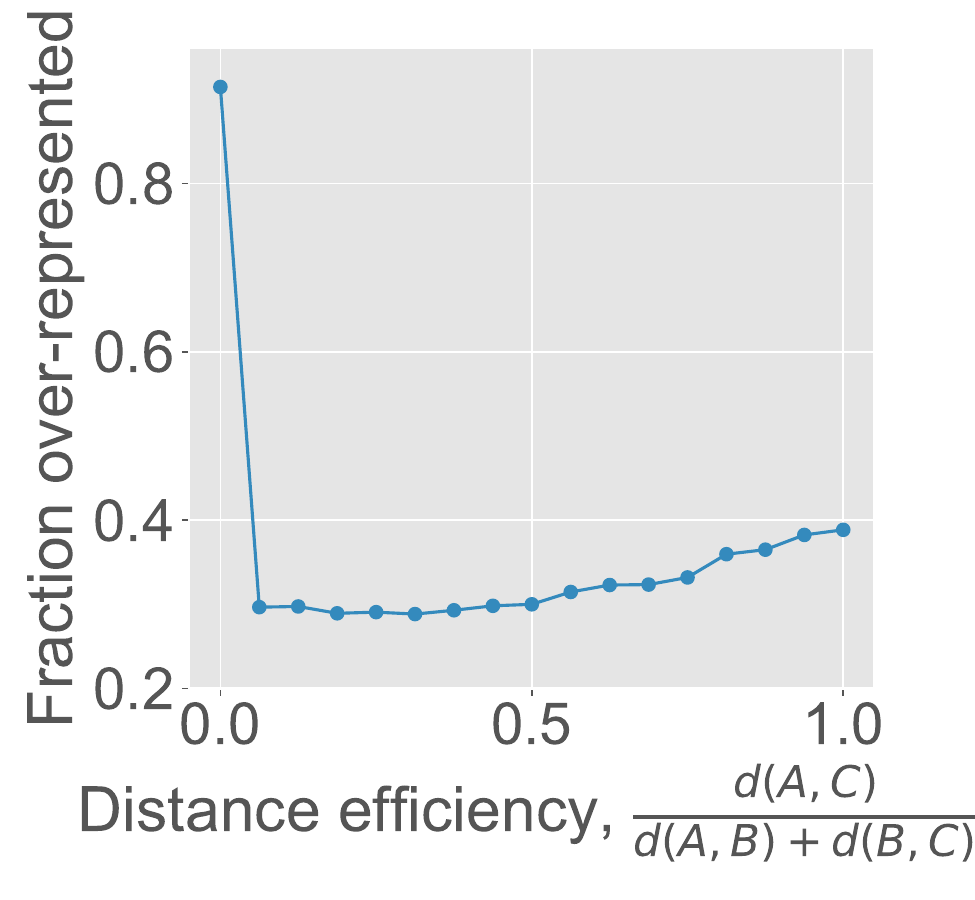}\hfill
    \vspace{-0.3cm}
    \caption{\small Left: extreme values of \emph{balance} correspond to over-represented paths, confirming that short flights followed by long flights are typical (e.g. flights to a regional hub, then a national hub). Middle/Right: The fraction of over- and under-represented paths varies with the efficiency of the itinerary. After return flights are accounted for, the fraction of under-represented paths decreases with efficiency (middle), and vice versa for over-represented paths (right).}
    \label{fig:rel-dist-overrep}
\end{figure}

However, we still observe a number of over-represented non-return flight paths.
We hypothesize that many of these paths connect small airports to large airports via regional hubs. 
This means that a relatively short distance trip (e.g. from ORL to ATL) is required before a flight from the regional hub to a relatively distant destination (e.g. ATL to LAX).
Rather than classifying airports by their size and role in the network, we test this hypothesis by defining \emph{distance balance}, a measure that captures to what extent one leg of a trip dominates the total trip distance.
In a perfectly balanced trip $(ABC)$, the distance of the two legs is equal, e.g. $d(A,B)=d(B,C)$.
The most common example of a perfectly balanced trip is the return trip, where $A=C$.
In an imbalanced trip, one of the legs of the trip is much larger.
We define balance by the ratio $\frac{d(A,B)-d(B,C)}{d(A,B) + d(B,C)}$.
It approaches -1 or 1 when the distance of one leg of the trip is much greater than on the other. 
We expect flights with extreme values to be over-represented as they represent long distance flights that start from small, local airports, fly a short distance to a regional hub, then on to a much further off destination (as well as the reverse). 
The distribution of balance for over- and under-represented paths of length two ($\alpha=0.05$) is shown in \cref{fig:rel-dist-overrep} (left).
We find very few under-represented flights near extreme values of balance, while a larger fraction of over-represented paths are found near -1 and 1.
This supports our hypothesis that unbalanced flights tend to be more over-represented than balanced flights.

We now formulate hypotheses based on a notion of \emph{efficiency} for airline trips. 
We measure efficiency as the ratio of the distance between source and destination, $d(A,C)$, with the actual flight distance, $d(A,B) + d(B,C)$.
Using this measure, a straight line between airports A, B and C has maximum efficiency of 1, while a low efficiency trip implies that the actual flight distance is much larger than the straight line distance between the origin and destination. 
We hypothesize that highly efficient paths are over-represented, while inefficient paths are under-represented in the data.
The middle plot of \cref{fig:rel-dist-overrep} shows a large peak in the fraction of under-represented paths at very low efficiency, then a steady decrease in under-represented paths as efficiency increases.
In the right hand plot we see that after return flights are accounted for (peak at efficiency 0), the fraction of over-represented paths increases monotonically with efficiency.
These results indicate that more efficient paths are indeed more likely to be over-represented, and that the more efficient a path is, the less likely it is to be under-represented.

\vspace{-0.25cm}
\section{Conclusion}
\label{sec:conclusion}

We presented \methodname, a novel approach for unsupervised detection of path anomalies in sequential data on networks.
By providing a new theoretical basis for anomaly detection in graphs, our work advances the state-of-the-art in multiple directions.
We introduced the problem of path anomaly detection and showed that frequency-based anomaly detection techniques cannot address it.
Projecting paths through a first-order network onto higher-order De Bruijn graphs, we showed that path anomaly detection can be reduced to the detection of anomalous edge weights in a higher-order graph.
Building on an analytically tractable null model of higher-order De Bruijn graphs, we developed a scalable method, \methodname\, that is able to detect paths that exhibit significant deviations from a random baseline, allowing us to assess statistical deviations in frequencies of paths traversing the nodes of a graph.
Some limitations of \methodname\ could be addressed in future work, including (i) automatically selecting an appropriate discrimination threshold, (ii) combining analysis at different orders, and (iii )incorporating domain specific notions of anomalous paths.

\paragraph{Acknowledgements}
\small IS acknowledges support by Swiss National Science Foundation grant 176938. LaRock and Eliassi-Rad were supported by (1) the Combat Capabilities Development Command Army Research Laboratory under Cooperative Agreement Number W911NF-13-2-0045 (ARL Cyber Security CRA) and (2) the Under Secretary of Defense for Research and Engineering under Air Force Contract No. FA8702-15-D-0001. The views and conclusions contained in this document are those of the authors and should not be interpreted as representing the official policies, either expressed or implied, of the Combat Capabilities Development Command Army Research Laboratory or the Under Secretary of Defense for Research and Engineering or the U.S. Government. The U.S. Government is authorized to reproduce and distribute reprints for Government purposes not withstanding any copyright notation here on.
\vspace{-0.4cm}
\bibliographystyle{abbrv}
\footnotesize\bibliography{library}

\begin{thebibliography}{1}

\bibitem{transstats}
{Bureau of Transportation Statistics}.
\newblock Transstats {OST\_R} database.
\newblock \url{http://www.transtats.bts.gov/Tables.asp?DB_ID=125}, 2019.

\bibitem{INSPIRE}
INSPIRE.
\newblock Inspire hep.
\newblock \url{http://inspirehep.net/info/hep/api}, 2018.

\bibitem{Merikoski1984}
J.~K. Merikoski.
\newblock On the trace and the sum of elements of a matrix.
\newblock {\em Linear Algebra and its Applications}, 60:177 -- 185, 1984.

\bibitem{TFL}
{Transport for London}.
\newblock Rolling origin and destination survey (rods)database.
\newblock \url{http://www.tfl.gov.uk/info-for/open-data-users/our-feeds}, 2014.

\bibitem{HOSPITAL}
P.~Vanhems, A.~Barrat, C.~Cattuto, J.~Pinton, and N.~Khanafer.
\newblock Estimating potential infection transmission routes in hospital wards
  using wearable proximity sensors.
\newblock {\em PLoS ONE}, 8(9):e73970, 09 2013.

\bibitem{West2012}
R.~West and J.~Leskovec.
\newblock Human wayfinding in information networks.
\newblock In {\em WWW}, pages 619--628, 2012.

\end{thebibliography}


\begin{thebibliography}{10}

\bibitem{agrawal1995mining}
R.~Agrawal and R.~Srikant.
\newblock Mining sequential patterns.
\newblock In {\em IEEE ICDE}, pages 3--14, 1995.

\bibitem{akoglu2010event}
L.~Akoglu and C.~Faloutsos.
\newblock Event detection in time series of mobile communication graphs.
\newblock In {\em Army Science Conference}, pages 77--79, 2010.

\bibitem{akoglu2015graph}
L.~Akoglu, H.~Tong, and D.~Koutra.
\newblock Graph based anomaly detection and description: a survey.
\newblock {\em Data Mining and Knowledge Discovery}, 29(3):626--688, 2015.

\bibitem{atzmueller2016detecting}
M.~Atzmueller.
\newblock Detecting community patterns capturing exceptional link trails.
\newblock In {\em ASONAM}, pages 757--764, 2016.

\bibitem{atzmueller2016sequential}
M.~Atzmueller, A.~Schmidt, and D.~Arnu.
\newblock Sequential modeling and structural anomaly analytics in industrial
  production environments.
\newblock In {\em LWDA}, pages 283--290, 2016.

\bibitem{becker2017mixedtrails}
M.~Becker, F.~Lemmerich, P.~Singer, M.~Strohmaier, and A.~Hotho.
\newblock Mixedtrails: Bayesian hypothesis comparison on heterogeneous
  sequential data.
\newblock {\em Data Mining and Knowledge Discovery}, 31(5):1359--1390, 2017.

\bibitem{bertens2016keeping}
R.~Bertens, J.~Vreeken, and A.~Siebes.
\newblock Keeping it short and simple: Summarising complex event sequences with
  multivariate patterns.
\newblock In {\em KDD}, page 735–744, 2016.

\bibitem{boden2012tracing}
B.~Boden, S.~G{\"u}nnemann, and T.~Seidl.
\newblock Tracing clusters in evolving graphs with node attributes.
\newblock In {\em CIKM}, pages 2331--2334, 2012.

\bibitem{bogdanov2011mining}
P.~Bogdanov, M.~Mongiov{\`i}, and A.~K. Singh.
\newblock Mining {{Heavy Subgraphs}} in {{Time}}-{{Evolving Networks}}.
\newblock In {\em IEEE ICDM}, pages 81--90, 2011.

\bibitem{cadez2000visualization}
I.~Cadez, D.~Heckerman, C.~Meek, P.~Smyth, and S.~White.
\newblock Visualization of navigation patterns on a web site using model-based
  clustering.
\newblock In {\em KDD}, pages 280--284, 2000.

\bibitem{casiraghi2018generalised}
Casiraghi and Nanumyan.
\newblock Generalised hypergeometric ensembles of random graphs: The
  configuration model as an urn problem.
\newblock {\em arXiv:1810.06495 [physics]}, 2018.

\bibitem{Casiraghi2017}
G.~Casiraghi, V.~Nanumyan, I.~Scholtes, and F.~Schweitzer.
\newblock From relational data to graphs: Inferring significant links using
  generalized hypergeometric ensembles.
\newblock In {\em Social Informatics}, pages 111--120, 2017.

\bibitem{chakrabarti1998mining}
S.~Chakrabarti, S.~Sarawagi, and B.~Dom.
\newblock Mining surprising patterns using temporal description length.
\newblock In {\em VLDB}, volume~98, pages 606--617, 1998.

\bibitem{chandola2009anomaly}
V.~Chandola, A.~Banerjee, and V.~Kumar.
\newblock Anomaly detection: A survey.
\newblock {\em ACM computing surveys}, 41(3):15, 2009.

\bibitem{chandola2012}
V.~Chandola, A.~Banerjee, and V.~Kumar.
\newblock Anomaly detection for discrete sequences: A survey.
\newblock {\em IEEE TKDE}, 24(5):823--839, 2012.

\bibitem{chandola2008comparative}
V.~Chandola, V.~Mithal, and V.~Kumar.
\newblock Comparative evaluation of anomaly detection techniques for sequence
  data.
\newblock In {\em IEEE ICDM}, pages 743--748, 2008.

\bibitem{DeBruijn1946}
N.~G. de~Bruijn.
\newblock A combinatorial problem.
\newblock {\em Koninklijke Nederlandse Akademie v. Wetenschappen}, 49:758--764,
  1946.

\bibitem{Sayed2004}
M.~El-Sayed, C.~Ruiz, and E.~A. Rundensteiner.
\newblock Fs-miner: efficient and incremental mining of frequent sequence
  patterns in web logs.
\newblock In {\em WIDM}, pages 128--135, 2004.

\bibitem{erdos1960evolution}
P.~Erdos and A.~R{\'e}nyi.
\newblock On the evolution of random graphs.
\newblock {\em Publ. Math. Inst. Hung. Acad. Sci}, 5(1):17--60, 1960.

\bibitem{gilbert1959random}
E.~N. Gilbert.
\newblock Random graphs.
\newblock {\em The Annals of Mathematical Statistics}, 30(4):1141--1144, 1959.

\bibitem{gupta2014}
M.~Gupta, J.~Gao, C.~C. Aggarwal, and J.~Han.
\newblock Outlier detection for temporal data: A survey.
\newblock {\em TKDE}, 26(9):2250--2267, 2014.

\bibitem{gwadera2005markov}
R.~Gwadera, M.~Atallah, and W.~Szpankowski.
\newblock Markov models for identification of significant episodes.
\newblock In {\em SDM}, pages 404--414, 2005.

\bibitem{gwadera2010ranking}
R.~Gwadera and F.~Crestani.
\newblock Ranking {{Sequential Patterns}} with {{Respect}} to {{Significance}}.
\newblock In {\em Advances in {{Knowledge Discovery}} and {{Data Mining}}},
  pages 286--299. 2010.

\bibitem{jianpei2001prefixspan}
{Jian Pei}, {Jiawei Han}, B.~{Mortazavi-Asl}, H.~Pinto, {Qiming Chen},
  U.~Dayal, and {Mei-Chun Hsu}.
\newblock {{PrefixSpan}},: Mining sequential patterns efficiently by
  prefix-projected pattern growth.
\newblock In {\em ICDE}, pages 215--224, 2001.

\bibitem{keogh2007compression}
E.~Keogh, S.~Lonardi, C.~A. Ratanamahatana, L.~Wei, S.-H. Lee, and J.~Handley.
\newblock Compression-based data mining of sequential data.
\newblock {\em Data Mining and Knowledge Discovery}, 14(1):99--129, 2007.

\bibitem{lane2003empirical}
T.~Lane and C.~E. Brodley.
\newblock An empirical study of two approaches to sequence learning for anomaly
  detection.
\newblock {\em MLJ}, 51(1):73--107, 2003.

\bibitem{larock2019detecting}
T.~LaRock, V.~Nanumyan, I.~Scholtes, G.~Casiraghi, T.~{Eliassi-Rad}, and
  F.~Schweitzer.
\newblock Hypa: Efficient detection of path anomalies in time series data on
  networks.
\newblock {\em arXiv preprint arXiv:1905.10580 [physics]}, 2020.

\bibitem{laxhammar2014online}
R.~Laxhammar and G.~Falkman.
\newblock Online learning and sequential anomaly detection in trajectories.
\newblock {\em TPAMI}, 36(6):1158--1173, 2014.

\bibitem{lemmerich2016mining}
F.~Lemmerich, M.~Becker, P.~Singer, D.~Helic, A.~Hotho, and M.~Strohmaier.
\newblock Mining subgroups with exceptional transition behavior.
\newblock In {\em KDD}, pages 965--974, 2016.

\bibitem{liu2018subgraph}
Z.~Liu, V.~W. Zheng, Z.~Zhao, H.~Yang, K.~C.-C. Chang, M.~Wu, and J.~Ying.
\newblock Subgraph-augmented path embedding for semantic user search on
  heterogeneous social network.
\newblock In {\em WWW}, pages 1613--1622, 2018.

\bibitem{melnyk2016vector}
I.~Melnyk, B.~Matthews, H.~Valizadegan, A.~Banerjee, and N.~Oza.
\newblock Vector autoregressive model-based anomaly detection in aviation
  systems.
\newblock {\em J. of Aerospace Information Systems}, 13:161--173, 2016.

\bibitem{Molloy1995}
M.~Molloy and B.~Reed.
\newblock {A critical point for random graphs with a given degree sequence}.
\newblock {\em Random Structures {\&} Algorithms}, 6(2-3):161--180, 1995.

\bibitem{Noble2003}
C.~C. Noble and D.~J. Cook.
\newblock Graph-based anomaly detection.
\newblock In {\em KDD}, pages 631--636, 2003.

\bibitem{peixoto2017modelling}
T.~P. Peixoto and M.~Rosvall.
\newblock Modelling sequences and temporal networks with dynamic community
  structures.
\newblock {\em Nature communications}, 8(1):582, 2017.

\bibitem{Rosvall2014}
M.~Rosvall, A.~V. Esquivel, A.~Lancichinetti, J.~D. West, and R.~Lambiotte.
\newblock Memory in network flows and its effects on spreading dynamics and
  community detection.
\newblock {\em Nature communications}, 5, 2014.

\bibitem{sadoddin2016finding}
R.~Sadoddin, J.~Sander, and D.~Rafiei.
\newblock Finding {{Surprisingly Frequent Patterns}} of {{Variable Lengths}} in
  {{Sequence Data}}.
\newblock In {\em SDM}, pages 27--35, 2016.

\bibitem{Scholtes2017}
I.~Scholtes.
\newblock When is a network a network?: Multi-order graphical model selection
  in pathways and temporal networks.
\newblock In {\em KDD}, pages 1037--1046, 2017.

\bibitem{servan2018prosecco}
S.~Servan-Schreiber, M.~Riondato, and E.~Zgraggen.
\newblock Prosecco: Progressive sequence mining with convergence guarantees.
\newblock In {\em IEEE ICDM}, pages 417--426, 2018.

\bibitem{singer2015}
P.~Singer, D.~Helic, A.~Hotho, and M.~Strohmaier.
\newblock Hyptrails: A bayesian approach for comparing hypotheses about human
  trails on the web.
\newblock In {\em WWW}, pages 1003--1013, 2015.

\bibitem{smyth1997clustering}
P.~Smyth.
\newblock Clustering sequences with hidden markov models.
\newblock In {\em NIPS}, pages 648--654, 1997.

\bibitem{tajer2014outlying}
A.~Tajer, V.~V. Veeravalli, and H.~V. Poor.
\newblock Outlying sequence detection in large data sets: A data-driven
  approach.
\newblock {\em IEEE Signal Processing Magazine}, 31(5):44--56, 2014.

\bibitem{tonnelier2018anomaly}
E.~Tonnelier, N.~Baskiotis, V.~Guigue, and P.~Gallinari.
\newblock Anomaly detection in smart card logs and distant evaluation with
  twitter: a robust framework.
\newblock {\em Neurocomputing}, 298:109--121, 2018.

\bibitem{tonon2019permutation}
A.~Tonon and F.~Vandin.
\newblock Permutation {{Strategies}} for {{Mining Significant Sequential
  Patterns}}.
\newblock In {\em IEEE ICDM}, 2019.

\bibitem{walk2014}
S.~Walk, P.~Singer, and M.~Strohmaier.
\newblock Sequential action patterns in collaborative ontology-engineering
  projects: A case-study in the biomedical domain.
\newblock In {\em CIKM}, pages 1349--1358, 2014.

\bibitem{zhou2016bi}
D.~Zhou, J.~He, Y.~Cao, and J.-S. Seo.
\newblock Bi-level rare temporal pattern detection.
\newblock In {\em IEEE ICDM}, 2016.

\end{thebibliography}

\clearpage

\appendix
\raggedbottom
\section{Supplementary Material}
To facilitate reproducibility and for completeness, we use this appendix to: 
(i) show some further analyses using HYPA that could not fit in the main text;
(ii) provide detailed pseudocode for \methodname and a proof of its computational complexity;  
(iii) provide pseudocode for the naive baseline method FBAD;  
(iv) explain in detail how the results in \cref{fig:auroc} were generated, including details and pseudocode for the synthetic model used to generate the data;
(v) describe in more detail the procedure we used to generate ground truth path anomalies and validate that \methodname\ discovers the expected anomalies in real data;
(vi) show the scalability of the algorithm on real data;   
and finally (vii) provide details of the construction and preprocessing of the real datasets.

\subsection{Further Analysis}
Here we present two further analyses using \methodname.

\subsubsection{Tube Geographic Hypothesis}
We use \methodname\ to test a hypothesis about the relationship between over- and under-representation of paths in the  \emph{Tube} dataset and the geographic distance the paths cover.
Intuitively, we expect people to use public transportation like the London Underground preferentially for longer distance trips, such as commutes to and from work, while avoiding trips with very short, walk-able distances.
This leads to the hypothesis that over-represented itineraries span larger geographic distances compared to those that are under-represented.

\begin{figure}[h]
\centering
\includegraphics[width=0.7\linewidth]{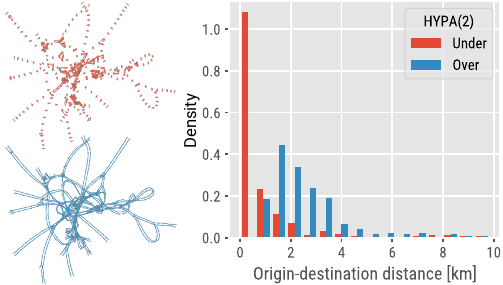}
\caption{\small \methodname acts as a filter on the $k$th order De Bruijn graph, separating over-represented (blue) and under-represented (red) paths of length \(k\). 
In the Tube data, under-represented paths detected by \methodname are tightly clustered, corresponding to avoidance of geographically short trips.
Over-represented paths appear in chains, corresponding to longer distance trips (as in daily suburban commutes). 
}%
\label{fig:london-tube}
\end{figure}

We compute \(\text{\methodname}^{(k)}\) scores for values $k=1, \ldots, 6$.
We then detect over- and under-represented paths based on discrimination threshold $\alpha = 0.001$.
We can use the detected anomalies to generate a decomposition of a $k$-th order De Bruijn graph model of paths based on under- and over-represented edge weights.
The resulting decomposition for $k=2$ are shown in \cref{fig:london-tube} (left column), where the nodes are placed according to geographic positions of London Tube stations.
The network of under-represented paths (top) exhibits high clustering and an absence of long chains, highlighting that it is predominantly paths spanning short geographic distances that are under-represented.
In contrast, the network of over-represented paths (bottom) shows long chains, supporting the hypothesis that paths spanning longer distances occur more often than expected at random.

To substantiate this intuition, in \cref{fig:london-tube} we show how geographic distances between start and end stations in over- and under-represen\-ted itineraries at $k=2$ are distributed.
The distance distribution of under-represented itineraries is shifted towards smaller distances, while the over-represented distribution is shifted towards larger distances, which supports our hypothesis. 
We find that the shift in distance distribution at order two is significant, witnessed by a $p$-value$\approx 0$ of a Mann-Whitney U-test (\cref{tab:tube-distances}).
Because the null model allows for paths starting and ending at the same node, there is a peak in the distribution of under-represented paths (see \cref{fig:london-tube}) at distance zero due to such paths being absent in the data.

At \(k>2\) paths starting and ending at the same node are not present, hence repeating the test provides stronger support for the geographic hypothesis.
As shown in \cref{tab:tube-distances}, the over-represented trips are indeed longer on average.

\begin{table}
    \caption{\small Median distance in kilometers between origins and destinations is significantly larger for over-represented paths of length \(k\) compared to under-represented paths in the London Tube, as shown by the \(p\)-value of a one-sided Mann-Whitney U-test.
    The discrimination threshold on \methodname\!(2) scores was $\alpha=0.001$.}
    \label{tab:tube-distances}
    \resizebox{0.9\linewidth}{!}{
    \begin{tabular}{@{}rrrrrrr@{}}
        \textbf{\(\text{\methodname}^{(k)}\)}    &   \(k=2\)  &   \(k=3\)  &   \(k=4\)  &   \(k=5\)  &   \(k=6\)  \\
        \hline    
        \textbf{Under [km]}        &   0.00   &   2.38   &   3.29   &   4.60   &   5.43   \\
        \textbf{Over [km]}         &   2.20   &   2.93   &   3.79   &   5.21   &   5.63   \\
        \hline
        \textbf{\(p\)-value}   & \(<\num{e-170}\) &  \(< \num{e-7}\)  &  \(< \num{e-4}\)  &  0.006  &  0.08  \\
    \end{tabular}
    }
\end{table}

\subsubsection{Path motifs}
\begin{figure}[ht]
    \includegraphics[width=0.9\linewidth]{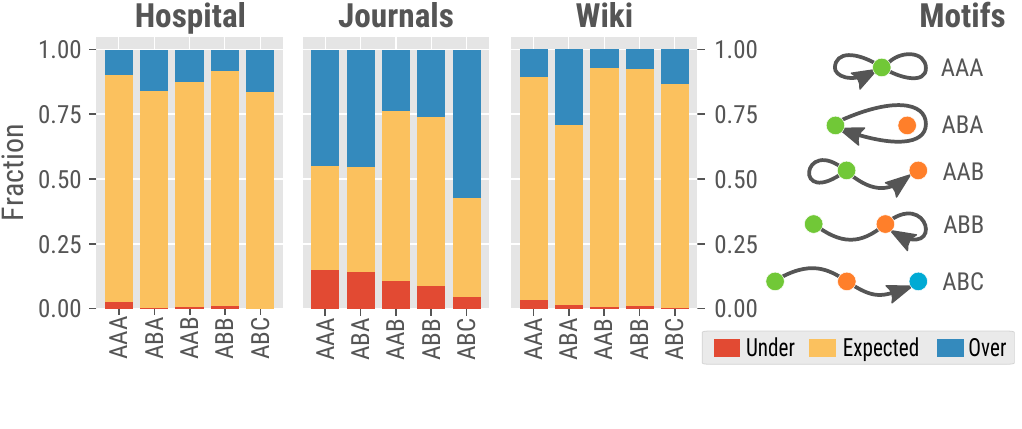}
    \caption{The distribution of over and under represented paths ($k=2$) across motifs is system dependent.
    The discrimination threshold used for detection was \(0.01\).
    }\label{fig:motifs}
\end{figure}

In \cref{fig:motifs} we study how anomalous paths detected by \methodname\ at \(k=2\) are distributed in the three data sets \emph{Wiki}, \emph{Journals}, and \emph{Hospital}.
We focus on five distinct motifs (horizontal axis in plots) where, e.g. \(\vv{ABC}\) represents paths traversing distinct nodes and \(\vv{ABA}\) represents paths that start and end in the same node but pass through another node.
While \cref{fig:motifs} highlights the absence of a universal pattern of motif anomalies across systems, some of the observed differences can be intuitively attributed to system-specific mechanisms.
For instance, paths of the type \(\vv{ABA}\) are among the most over-represented paths in \emph{Wiki}, which is likely a result of users using the `back' button of their browser while playing the game.
In \emph{Journals}, citation paths of the type \(\vv{ABC}\) through three distinct journals are both most over-represented and least under-represented.
This indicates (i) a hierarchy in journals in terms of knowledge flows through citations (papers in \(A\) implicitly rely on papers in \(C\)) and (ii) that knowledge flow is preferentially routed through certain sets of (probably multi-disciplinary) journals \(B\).

\subsection{\methodname\ Pseudocode and Details}
\label{app:HYPA}
In this section we present the pseudocode for both the \methodname\ algorithm and the algorithm to redistribute the values of the \(\Xi\) matrix, as well as a proof of the computational complexity of \methodname.
\begin{algorithm}[h]\small
	\caption{ComputeHYPA\((\mat{S},k)\): \emph{Compute \(k\)th order HYPA scores for sequence dataset \(S\)}.}
	\begin{algorithmic}[1]
		\REQUIRE \(\mat{S}\) (sequences), \(k\) (desired order)
		\ENSURE  \(\textrm{\methodname\!}^{(k)} \) score for all \(k\)-th order paths
		\STATE \(G^k {\gets}\)DeBruijnGraph(\(\mat{S}\),\(k\)) \COMMENT{Construct \(k\)th order graph}
		\STATE \(\Xi {\gets}\) fitXi(\(G^k\), tolerance)
		\FOR {$(\vv{v}, \vv{w}) \in G^k$}	
			\STATE \(\text{\methodname}^{(k)}(\vv{v}, \vv{w}) {\gets} \Pr(x_{vw} {\leq} (\vv{v}, \vv{w}) \,|\, m, \Xi) \) \COMMENT{Compute CDF}
		\ENDFOR
		\RETURN \(\text{\methodname}^{(k)}\) 
\end{algorithmic}
\label{alg:hypa_algo}
\end{algorithm}

\subsubsection{$\Xi$ Redistribution}%
\label{app:fitXi}
In \cref{sec:method:hypa} we briefly describe a simple algorithm to redistribute the values in the matrix \(\Xi\) such that it respects the constraints imposed by the \(k\)-order De Bruijn graph while also preserving the weighted in- and out-frequencies of the \(k\)-order nodes when used for sampling.
\Cref{alg:em_algo} shows the exact procedure we employ.

\begin{algorithm}[h]\small
\caption{fitXi(\(G^k\), tolerance): \emph{Adjust entries of \(\Xi\) to match expected frequencies (based on \(\Xi\)) to observed frequencies (based on \(W\))within ``tolerance'' error.}}\label{alg:fitxi}
\begin{algorithmic}[1]
	\REQUIRE \(G^k\) (\(k\)-order De Bruijn graph),  tolerance
	\ENSURE \(\Xi\)
    \STATE \(f_v^{\text{out}} {\gets} \sum_{x}f(v,x)\)    \COMMENT{weighted out-degrees}
    \STATE \(f_v^{\text{in}} {\gets} \sum_{x}f(x,v)\)     \COMMENT{weighted in-degrees}
    \STATE \(m {\gets} \sum_{v} f_v^{\text{out}}\)        \COMMENT{sum of all weights}
    \STATE \(\Xi_{vw} {\gets} f_v^{\text{out}} \cdot f_w^{\text{in}}\)  \COMMENT{initialize matrix for all \(v\),\(w \in G^k\)}     
    \STATE \(M {\gets} \sum_{vw} \Xi_{vw}\) 
    \IF{edge \((v,w)\) not possible in \(G^k\)}
        \STATE \(\Xi_{vw} {\gets} 0\)         \COMMENT{ensure edge \((v,w)\) cannot be sampled}
    \ENDIF
    \REPEAT
		\STATE \(\widehat f_v^{\text{in}} {\gets}  \frac{\sum_x \Xi_{xv}}{m}\frac{M}{\sum_{vw} \Xi_{vw}}\)    \COMMENT{Expectation for in-degrees}
		\STATE \(\Xi_{vw} {\gets} \Xi_{vw} \cdot \frac{f_w^{\text{in}}}{\widehat f_w^{\text{in}}}\)  \COMMENT{Correction for in-degrees}
        \STATE \(\widehat f_v^{\text{out}} {\gets}  \frac{\sum_x \Xi_{vx}}{m}\frac{M}{\sum_{vw} \Xi_{vw}}\)    \COMMENT{Expectation for out-degrees}
		\STATE \(\Xi_{vw} {\gets} \Xi_{vw} \cdot \frac{f_v^{\text{out}}}{\widehat f_v^{\text{out}}}\) \COMMENT{Correction for out-degrees}
	\UNTIL \(\textrm{RMSE}(f^{\text{out}}, \widehat f^{\text{out}}) {+} \textrm{RMSE}(f^{\text{in}}, \widehat f^{\text{in}}) {\leq}\) tolerance 
	\RETURN \(\Xi\)
\end{algorithmic}
\label{alg:em_algo}
\end{algorithm}

\subsubsection{Proof of Computational Complexity}
\label{sec:proof-complexity}
As stated in \cref{sec:complexity}, the asymptotic runtime of HYPA is $\mathcal{O}\left(N+\Delta^k(G)\right)$, where $\Delta^k(G)$ is the number of edges in a $k$-th order De Bruijn graph model $G^k$ of paths in $G$. In this section, we prove an upper bound on $\Delta^k(G)$:
{
\setcounter{lemma}{0}
\begin{lemma}
Let $G=(V,E)$ be a directed graph and let $k \in \mathbb{N}$ be the order of a De Bruijn graph model of paths in $G$. $\Delta^k(G)$ is bounded above by $|V|^2\lambda_1^k$, where $\lambda_1$ is the leading eigenvalue of the binary adjacency matrix of $G$.
\end{lemma}
}

\begin{proof}
We first note that for a fully connected graph $G$ with $|V|$ nodes and $E=V^2$ we trivially have $\Delta^k(G)=|V|^{k+1}$.
This follows from the fact that every possible sequence of $k+1$ nodes is a path of length $k$ in a full graph, i.e. a $k$-th order De Bruijn graph model has $|V|^{k+1}$ edges.

Let us now consider a graph $G$ with arbitrary topology and let $\mathbf{A}$ be the binary adjacency matrix of $G$, where one-elements indicate the presence and zero-elements indicate the absence of an edge.
We can compute the number of distinct paths of length $k$ in $G$ as \( \sum_{i}\sum_{j} (\mathbf{A}^k)_{ij} \), where $\mathbf{A}^k$ is the $k$-th power of adjacency matrix $\mathbf{A}$.
This directly follows from the definition of matrix multiplication, leading to the fact that each element $(\mathbf{A}^k)_{ij}$ in the $k$-th power of $\mathbf{A}$ counts distinct paths of length $k$ between nodes $i$ and $j$.

To prove the lemma, we use the following two facts.
First, the sum of all elements in any matrix $\mathbf{B}$ is equal to $\text{tr}(\mathbf{J}\mathbf{B})$, i.e., the sum of \emph{diagonal} elements in the matrix product $\mathbf{J}\mathbf{B}$, where $\mathbf{J}$ is the $|V|\times|V|$ all-ones matrix.
Second, if $\mathbf{B},\,\mathbf{C}$ are positive semi-definite matrices, from the Cauchy-Schwartz inequality follows $\text{tr}(\mathbf{C}\mathbf{B}) \leq \text{tr}(\mathbf{C})\text{tr}(\mathbf{B})$~\citeappendix{Merikoski1984}.
We can thus write:

$$ \Delta^k(G) = \sum_{i=1}^{|V|}\sum_{j=1}^{|V|} (\mathbf{A}^k)_{ij} = \text{tr}(\mathbf{J}\mathbf{A}^k) \leq \text{tr}(\mathbf{J}) \cdot \text{tr}(\mathbf{A}^k) $$

We now recall (i) that the trace of any square matrix is equal to the sum of its eigenvalues, (ii) that the eigenvalue sequence of an $|V| \times |V|$ all-one matrix $\mathbf{J}$ is $|V|, 0, \ldots, 0$, and (iii) that the eigenvalues of the $k$-th power $\mathbf{A}^k$ of a matrix are the $k$-th powers of eigenvalues of $\mathbf{A}$.
We can thus write:

\[ \Delta^k(G) \leq \text{tr}(\mathbf{J}) \cdot \text{tr}(\mathbf{A}^k) = |V| \cdot \sum_{i=1}^{|V|}\lambda_i^k \]

where $\lambda_i$ are the (not necessarily unique) eigenvalues of $\mathbf{A}$.
%
%
Without loss of generality, we assume that the eigenvalues are given in descending order, i.e. $\lambda_1 \geq \ldots \geq \lambda_{|V|}$. 
Hence, an upper bound $\Delta^k(G) \leq |V|^2\lambda_1^k$ can be derived based on the largest eigenvalue $\lambda_1$ of the adjacency matrix of $G$.
We note that for the special case of a fully connected graph, where $\mathbf{A}=\mathbf{J}$, we have $\Delta^k(G) = \text{tr}(\mathbf{J}\mathbf{A}^k) = \text{tr}(\mathbf{J}^{k+1}) = \lambda_1^{k+1} = |V|^{k+1}$ and we thus recover the trivial case from above.
\end{proof}

\subsection{Frequency Based Anomaly Detection}
\label{app:naive}
See \cref{alg:naive} and \cref{sec:baseline}.
\begin{algorithm}[!h]\small
	\caption{FBAD($S, k, \alpha$): \emph{Given path data $S$, order $k$, and scaling factor $\alpha \in \mathbb{R}$, compute anomalies based on the distribution of order-$k$ edgeweights.}}
	\label{alg:naive}
	\begin{algorithmic}[1]
		\REQUIRE $S$ (input path data), $k$ (desired anomaly order), $\alpha$ (scaling factor)
		\ENSURE $G$ ($k$-th order network with anomaly-labeled transitions)
		\STATE $G\gets$DeBruijnGraph($S,k$)
		\STATE $\mu \gets$ average of edge weights in $G$
		\STATE $\sigma\gets$ standard deviation of edge weights in $G$
		\FOR {edge $e$ in $G$}
		\IF {frequency$(e)> \mu+\sigma\alpha$}
		\STATE Label $e$ over-represented
		\ELSIF {frequency$(e) < \mu-\sigma\alpha$}
		\STATE Label $e$ under-represented
		
		\ENDIF
		\ENDFOR
		\RETURN Labeled graph $G$
	\end{algorithmic}
	
\end{algorithm}

\subsection{ROC Curves used to compute AUC}
In \cref{fig:auroc,fig:groundtruth}, we presented area under the curve results for a binary classification experiment where we used \methodname to predict ground truth over- and under-represented paths. 
In this section, we clarify the procedure for generating these results.

First, we use Algorithms \ref{alg:synthetic-model} and \ref{alg:synthetic-walk} to generate a dataset with injected anomalies at order $l$. 
Then, for each value of $k$, we compute $\textrm{\methodname}^{(k)}$ scores,
and for increasing $\alpha$ from 0 (nothing detected) to 1 (everything predicted as significant), we threshold the $\textrm{\methodname}^{(k)}$ scores to classify the ground truth over- and under-represented paths. 
We compute the true and false positive rates for each $\alpha$, which results in a single ROC, where each point is a combination of $k$ and $\alpha$.
We get an ROC for every combination of \methodname order ($k$) and anomaly order ($l$).
Finally, we compute the area under these curves and report averages and standard deviations over many randomly generated datasets.

\subsection{Synthetic Data and Simulations}

\subsubsection{Synthetic model}
\label{app:synthetic}
In \cref{sec:data}, we briefly described a model that generates pathways with injected correlation patterns, but could not include the pseudocode due to space constraints. \cref{alg:synthetic-model} presents pseudocode for constructing the graph topology, and \cref{alg:synthetic-walk} shows how we generate a walk on this topology.

\begin{algorithm}[h]\small
	\caption{SyntheticModel($N, p, f, k$): \emph{Generates a directed $G_{N,p}$ graph and marks fraction $f$ of length $k$ pathways through $G$ anomalous.}}
	\label{alg:synthetic-model}
	\begin{algorithmic}[1]
		\REQUIRE $N$ (number of nodes), $p$ (connection probability), $f$ (fraction of anomalous pathways), $k$ (anomaly order)
		\ENSURE $G$ (weighted topology), paths (paths marked anomalous)
		\STATE $G \gets$ directedER(N, p)
		\FOR {$(i,j) \in G$}
		\STATE $G_{i,j} \gets$ unif(1, 20) \COMMENT{Assign edge weight}
		\ENDFOR
		\STATE $G^k \gets $ DeBruijnGraph($G$, $k$)
		\STATE anom-paths=$\emptyset$
		\FOR {path $\in G^k$}
		\IF {$random()  < f$}
		\STATE anom-paths$\gets$path \COMMENT{mark path anomalous} 
		\ENDIF
		\ENDFOR
		\RETURN $G$, anom-paths
	\end{algorithmic}
\end{algorithm}
\vspace{-0.5cm}
\begin{algorithm}[!h]\small
\caption{SyntheticWalk($G, \textrm{paths}, l$): \emph{Given a weighted first-order topology $G$ and list of anomalous paths, generate a (potentially anomalous) random walk of length $l$.}}
\label{alg:synthetic-walk}
\begin{algorithmic}[1]
	\REQUIRE $G$ (weighted network topology), paths (paths through $G$ marked anomalous), l (length of walk)
	\ENSURE path
	\STATE $u\gets$ uniform random node
	\STATE path = $[u]$
	\WHILE {$j<l$}
		\IF {$u$ is on an anomalous path}
			\WHILE {$j < l$ and nodes remain on anomalous path}
				\STATE $v \gets$ next node on anomalous path
				\STATE Append $v$ to path
				\STATE $j\gets j+1$
			\ENDWHILE
		\ELSE
				\STATE $v {\gets} v \in N_u$ \COMMENT{$\Pr(u,v) \propto$ edge weight $G_{u,v}$}
				\STATE Append $v$ to path
				\STATE $j\gets j+1$
		\ENDIF
		\STATE $u$ = $v$
	\ENDWHILE
	\RETURN path
\end{algorithmic}
\end{algorithm}

\subsubsection{Simulation-based Ground Truth Labels}
\label{app:ground-truth}
\begin{algorithm}[!h]\small
	\caption{GroundTruth($S, k, \alpha, m$): \emph{Given path data $S$, order $k$, discrimination threshold $\alpha \in \mathbb{R}$, and number of samples $m$, compute ground truth path anomalies based by numerical simulations.}}
	\label{alg:groundtruth}
	\begin{algorithmic}[1]
		\REQUIRE $S$ (input path data), $k$ (desired order), $\alpha$ (discrimination threshold), $m$ (number of datasets to simulate)
		\ENSURE $G$ ($k$-th order network with anomaly-labeled transitions)
		\STATE $G^{k-1}\gets$DeBruijnGraph($S,k-1$)
		\STATE $G^k\gets$DeBruijnGraph($S,k$)
		\FOR {m simulations}
		\STATE $S_{\textrm{rnd}}\gets $ RandomWalks($G^{k-1}, S$)
		\STATE Append frequency of each $kth$-order transition $e \in G^k$ to list
		\ENDFOR
		
		\FOR {edge $e$ in $G^k$}
		\STATE Estimate multinomial of $e$'s edgeweight $w$ as $\frac{x_w}{\sum_k{x_w}}$
		\STATE Compute CDF of multinomial distribution
		\IF {CDF($e$) $\geq 1-\alpha$}
		\STATE Label $e$ over-represented
		\ELSIF {CDF($e$) $< \alpha$}
		\STATE Label $e$ under-represented
		\ENDIF
		\ENDFOR
		\RETURN Labeled graph $G^k$
	\end{algorithmic}
\end{algorithm}
We evaluate the performance of \methodname\ in section~\ref{sec:baseline} using numerical simulations to generate the expected empirical frequency distributions of all paths of length $k$ in an empirical data set.
We achieve this by randomizing an empirical data set via a large number of $(k-1)$-order random walks, where the length of random walks matches the lengths of paths in the data set.
For multiple randomizations of the data, we obtain expected frequency distributions of all paths of length $k$.
Since computing the difference between actual and expected frequency (as in \cref{fig:toy-examples}(c)) does not give us a notion of significance, we sample multiple randomized path datasets and maintain a list of observed frequencies for each transition.
We use these random observations to estimate a multinomial distribution for the weight on each transition.
We then compute the Cumulative Distribution Function of this estimated distribution and filter the transitions using a discrimination threshold in the same way as in \methodname.
Based on this filtering, we obtain a ground truth against which we can compare \methodname\ and FBAD.

We note that in this analysis the construction of ground truth is dependent on the estimate of the multinomial distribution on each pathway, which in turn depends on the number of random datasets sampled ($m$ in \cref{alg:groundtruth}). 
Due to the computational cost of simulating the more than 4 million random walks in the Tube dataset, in these experiments the estimated multinomial was constructed using only 2 observations.

Here we provide pseudocode for the construction of ground truth path anomalies given a pathway dataset. Note that the results in \cref{fig:groundtruth} are impacted by the sampling parameter $m$: as $m$ increases, the discovered ground truth will get more accurate and the higher AUC will be for \methodname.
An implementation of our numerical technique to infer ground truth path anomalies is available online at \url{github.com/tlarock/hypa}.

\subsubsection{Simulated Tube Anomalies}
\begin{figure}[!ht]
	\centering
    \includegraphics[width=0.75	\linewidth]{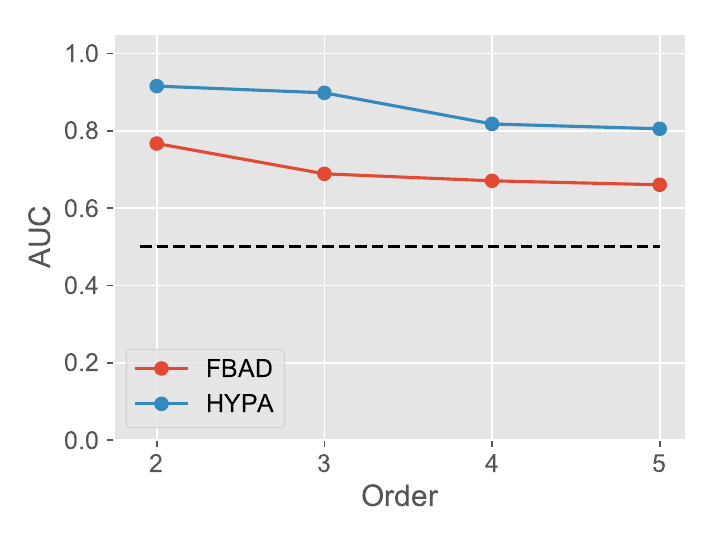}
    \caption{\small \methodname\ outperforms FBAD in detecting anomalous paths in the London Tube data. For this data set, ground truth anomalies can be established using computationally expensive numerical simulations.
    We apply both methods at various orders $k$ and measure their performance in predicting the ground truth.
    At all detection orders $k$, \methodname\ considerably outperforms FBAD, illustrating the inadequacy of frequency-based methods for path anomalies.
    }
    \label{fig:groundtruth}
\end{figure}

While simulating ground truth labels is prohibitively expensive for large data sets, it enables us to generate a proxy for ground truth path anomalies in the London Tube data.
In this case, each randomized version of the data is generated by performing more than 4.8 million random walks with an average length of $14.8$ steps in the (weighted) graph topology.
We repeat this multiple times to obtain ground truth labels for over- and under-represented paths.
Repeating the experiment from \cref{sec:synthetic}, these ground truth labels allow us to compare the performance of \methodname\ against the baseline frequency-based detection (FBAD).
We use the London Tube data set for our experiment because its topology is sufficiently small and sparse to allow for this expensive experiment.
The results in \cref{fig:groundtruth} show that (i) \methodname\ is able to detect ground truth path anomalies with high accuracy and (ii) our analytical approach outperforms the detection performance of the baseline frequency-based detection (FBAD) at all orders $k$.

\subsubsection{Scalability Analysis}
\begin{figure}[ht]
    \centering
    \includegraphics[width=.49\linewidth]{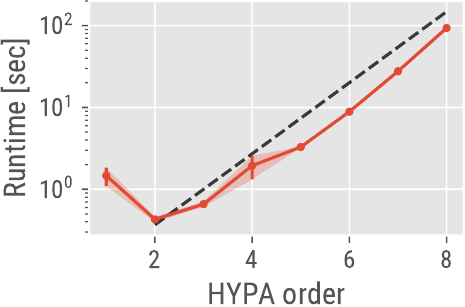}\hfill
    \includegraphics[width=.49\linewidth]{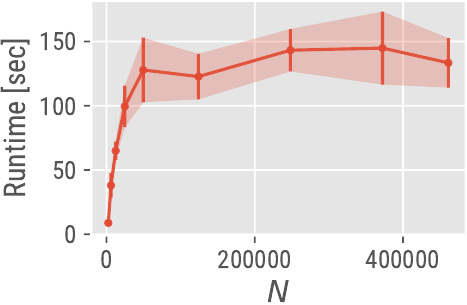}
    \caption{Empirical scalability of HYPA. Left: Required time to detect path anomalies of length $k$ for the Tube data. Right: Runtime in Flights data for detection order $k=1$ and varying data size $N$ randomly sampled from the data. All data points correspond to the mean of ten repeated measurements, with the standard deviations shown as bars.}
    \label{fig:scalability}
\end{figure}

We finally validate the theoretical analysis of computational complexity in  \cref{sec:complexity} through an experimental evaluation of scalability in empirical data.
We measure the time needed to detect path anomalies for (i) varying orders $k$ in a data set of fixed size $N$ and (ii) a fixed order $k$ and data with varying size $N$.
Fig.~\ref{fig:scalability} reports the time needed to run \methodname\ on a single core of an Intel i7-7600U CPU.
All values are averages of ten measurements.
The left panel in Fig.~\ref{fig:scalability} confirms that the runtime of our algorithm scales as an exponent of $k$, where the basis of the exponent depends on the algebraic connectivity of the graph.
We note however that, even for a data set with more than four million paths the detection of anomalies up to length eight takes less than two minutes.
The right panel in Fig.~\ref{fig:scalability} confirms that our analytical approach is suitable for large data sets.
In particular, the experimental results are aligned with our theoretical analysis of computational complexity in \cref{sec:method:hypa}, which predicts that below a critical sum of path lengths $N$, the runtime of \methodname\ is dominated by the number of paths of length $k$. 
This explains why for small values of $N$ we observe an exponential increase of runtime as the size of the $k$-dimensional De Bruijn graph model approaches the theoretical upper limit of $|V|^2\lambda_1^k$.
For large values of $N$, the runtime of HYPA is dominated by a linear term that is due to the single pass through the data, while the calculation of HYPA scores is independent of the size of the data.
This confirms that the analytical approach underlying our algorithm makes it suitable to analyze big time series data on networks.

\subsection{Data}
\label{app:data}
In \cref{tab:data-stats} we present some statistics about the real datasets. Below we provide more detail about processing of the data.
\vspace{0.25cm}
\begin{table}[h]
	\footnotesize
	\caption{\small Description of empirical data sets, with \(\lambda_1\) denoting the largest eigenvalue of the adjacency matrix of graph \(G\), \(N=\sum_i l_i\) being the sum of all path lengths, \(l^{\text{max}}\) and  \(\mean{l}\) denoting maximum and average path length.}
	\label{tab:data-stats}
	\resizebox{\linewidth}{!}{
		\begin{tabular}{@{}rrrrrrrrr@{}}
			& \multicolumn{3}{l}{\textbf{Graph \(G=(V,E)\)}} & \multicolumn{5}{l}{\textbf{Sequences \(S\)}} \\
			\hline
			Data         & |V| & |E| & \(\lambda_1\) &  Total   & Unique & \(l^{\text{max}}\) & \(\mean{l}\) & \(N\) \\
			\hline
			\textbf{Journals}   & 283 & 1743 & 26.19 &  480496  &  309565   &   35  &   14.8 & $1.64 \cdot 10^8$\\ 
			\textbf{Tube}       & 268 & 646  & 3.99 & 4295731  &   67015   &   35  &   6.75 & $2.89 \cdot 10^7$ \\ 
			\textbf{Flights}    & 382 & 6933 & 56.55 &  185871  &   88539   &   10  &   2.48 & $4.61 \cdot 10^5$ \\ 
			\textbf{Wiki}       & 100 & 1598 & 21.47 &   29682  &    7431   &   21  &   1.64 & $4.88 \cdot 10^4$ \\
			\textbf{Hospital}   & 75  & 1138 & 37.01 &   28422  &    2561   &    5  &   1.19 &  $3.37 \cdot 10^4$\\
			\hline
		\end{tabular}
	}
\end{table}

\paragraph{Flights} The flights dataset is given in the form of ``itineraries'', which correspond to tickets purchased together by a particular customer \citeappendix{transstats}.
Each pathway is a sequence of airports corresponding to source, layovers, and destination.
Our dataset is constructed by taking a uniform 5\% sample of these pathways from the first quarter of 2018.

\paragraph{Tube} The Tube data is given in the form of origin-destination statistics between stations \citeappendix{TFL}. 
We use these statistics in conjunction with the first-order topology of the Tube network to construct pathways by computing the shortest path between every station and assuming riders take this path. 
If there are multiple shortest paths between an origin and a destination, the observed paths are distributed across them.

\paragraph{Journal Citations} The journal citation data begins with a citation graph \citeappendix{INSPIRE}, where a directed link is drawn from paper $i$ to paper $j$ if $i$ cites $j$. 
We then enforce that this graph is directed and acyclic by removing ``backlinks'', meaning links from node $i$ to $j$ such that $j$ was published after $i$. 
Pathways of citations are then constructed by walking from a ``source'' paper (a paper which was never cited in the dataset) to a ``sink'' paper (a paper that didn't cite any other papers in the dataset).
Reversing the order of this pathway results in a chronological ``citation flow'' from the sink (the oldest paper) to the source (the newest paper).
These sequences of papers are then projected using a mapping from individual paper to publication venue, resulting in sequences of journals that cited one another through time. 

\paragraph{Hospital} The Sociopatterns data is a sequence of time stamped edges representing interactions between nurses, doctors, administrative staff and patients in a hospital \citeappendix{HOSPITAL}. 
We define a pathway by a 20-second inter-event time, meaning that if 2 interactions including a common person happened within 20 seconds, they are combined into a path. 
A path ends when 20 seconds passes without the last person to interact having a subsequent interaction.

\paragraph{Wikispeedia} We focus our analysis of the Wikispeedia data \citeappendix{West2012} on pathways representing finished games.
In the full dataset, the number of observed pathways is too small relative to the size and density of the underlying article graph to compute meaningful statistics.
Due to this, we only analyze games which traverse the 100 most frequently visited articles in the dataset, and filter out pathways of length less than 4.

\bibliographystyleappendix{abbrv}
\footnotesize\bibliographyappendix{library}

\end{document}